\documentstyle[psfig]{l-aa}

\begin{document}

   \thesaurus{05(08.05.3; 08.08.1; 08.12.3; 08.16.3; 10.07.3 NGC~1851 )          }
\title{The Galactic globular cluster NGC~1851: its dynamical and
evolutionary properties
\thanks{
Based on observations made at the European Southern Observatory, La Silla,
Chile, and archive {\it HST} observations retrieved through the {\sc starview}
interface.
Appendices A and B are only available in electronic form at
the CDS via anonymous ftp to cdsarc.u-strasbg.fr(130.79.128.5) or via
http://cdsweb.u-strasbg.fr/Abstract.html
 } } \subtitle{ }

   \author{I. Saviane$^1$, G. Piotto$^1$, F. Fagotto$^2$, S. Zaggia$^3$, 
           M. Capaccioli$^3$, A. Aparicio$^2$
          }

   \offprints{I. Saviane, e-mail: saviane@pd.astro.it }

   \institute{$^1$ Dipartimento di Astronomia -- Universit\`a di
Padova, Vicolo dell' Osservatorio, 5, I-35122 Padova, Italy \\
              $^2$ Instituto de Astrof\'{\i}sica de Canarias,
Via Lactea, E-38200 La Laguna, Tenerife, Spain \\
              $^3$ Osservatorio Astronomico di Capodimonte, Via Moiariello 16, I-80131 Napoli, Italy
            }

   \date{Received ; accepted }

   \maketitle

   \markboth{Saviane et al.}{VI photometry of NGC~1851}

   \begin{abstract}

We have completely mapped the Galactic globular cluster NGC~1851 with
large-field, ground-based $VI$ CCD photometry and pre-repair $HST$/WFPC1
data for the central region.  

The photometric data set has allowed a $V $ vs. $ (V-I) $ 
colour--magnitude diagram for $\sim$ 20500 stars to be constructed.
From the apparent luminosity of the horizontal branch (HB)
 we derive a true distance
modulus $(m-M)_0$ = 15.44 $\pm$ 0.20.

An accurate inspection of the cluster's bright and blue objects
confirms the presence of seven ``supra-HB'' stars, six of which are
identified as evolved descendants from HB progenitors.

The HB morphology is found to be clearly bimodal, showing both a red clump and a
blue tail, which are not compatible with standard evolutionary models.
Synthetic Hertzsprung--Russell (HR) diagrams demonstrate that 
the problem could be solved by
assuming a bimodal efficiency of the mass loss along the red giant branch (RGB).
  With the aid
of Kolmogorov--Smirnov statistics we find evidence that the radial
distribution of the blue HB stars is different from that of the red HB
and subgiant branch (SGB) stars.

We give the first measurement of the mean  absolute $I$ magnitude
for 22 known RR~Lyr variables ($<M_{I}({\rm RR})> = 0.12
\pm 0.20$ mag at a metallicity [Fe/H]~=~--1.28). The mean absolute $V$
magnitude is $<M_{V}(\rm RR)> = 0.58
\pm 0.20$ mag, and we
confirm that these stars are brighter than those of  the 
zero-age HB (ZAHB).  Moreover, we found seven
new RR~Lyr candidates (six $ab$ type and one $c$ type).
With these additional variables the ratio of the
two types is now $N_c$/$N_{ab} = 0.38$.

From a sample of 25 globular clusters a new calibration for $\Delta
V_{\rm bump}^{\rm HB}$
as a function of cluster metallicity is derived. 
NGC~1851 follows this general trend fairly well. From a comparison with
the theoretical models, we also find some evidence for an age--metallicity
relation among globular clusters.

We identify 13 blue straggler stars, which do not show any sign of
variability. The blue stragglers are less concentrated than the subgiant
branch stars with similar magnitudes for $r>80$ arcsec.

Finally, a radial dependence of the luminosity function, a sign of
mass segregation, is found.  Transforming the luminosity function into a mass
function (MF) and correcting for mass segregation by means of multi-mass
King--Michie models, we find a global MF exponent $x_0=0.2\pm
0.3$.

\end{abstract}

\section{Introduction}

Galactic globular clusters (GGC) are dynamically evolved objects.
In order to understand the interplay between the internal dynamical
processes and the influence of the Galactic potential, we must study a
sample of GGCs comprising objects whose concentration, position in the
Galaxy, luminosity and metallicity cover the whole observed range. The mass
function and the radial profile must be determined for each cluster, in
order to carry out a detailed dynamical analysis.

The introduction of large-size CCDs has made this kind of investigations
possible. With these detectors it is also possible to obtain deep photometry
for the nearest globulars, and therefore to probe their mass functions over
large mass intervals, in order to reach those MS stars which are more
sensitive to dynamical effects (e.g. Pryor et al. 1986).

A rich sample of stars is also essential in order to reveal and study the
shortest-lived (and hence poorly known) phases of the stellar evolution
(Renzini \& Buzzoni 1986). Furthermore, the interactions between the single
stars affect their evolution (e.g. Djorgovski et al. 1991).  To establish
the reliability of the stellar evolutionary models, we must therefore
ascertain to what extent a GC colour--magnitude diagram and luminosity
function is changed by the interactions among its stars.

For the above reasons, our group started a project aimed at studying a
number of globular clusters covering a wide range of the relevant
parameters.
NGC~1851 ($\alpha_{2000} = 5^{\rm h} 14^{\rm m} 6^{\rm s}.30$;
$\delta_{2000} = -40^\circ 2\arcmin 50\farcs00$) has been selected for its
position and its concentration. Its galactocentric
distance, which is about twice that of the Sun, and its distance of 7.1~kpc
from the Galactic plane (Djorgovski 1993)
 make it a typical halo object. Its concentration $c = 2.24$
is one of the highest in the list of Trager et al. (1995). A recent
measurement of the cluster's proper motion has  confirmed that
NGC~1851 has halo-type kinematics (Dinescu et al. 1996). According to these
authors, the space velocities of the cluster are $U=256\pm35$~km s$^{-1}$,
$V=-195\pm26$~km s$^{-1}$,  $W=-122\pm30$~km s$^{-1}$, $\Pi=195\pm37$~km s$^{-1}$ and
$\Theta=167\pm37$~km s$^{-1}$.

Past photometric studies of the cluster are given in Alcaino (1969, 1971,
1976), Stetson (1981, hereafter S81), Sagar et al. (1988), Alcaino et al.
(1990) and Walker (1992a, hereafter W92). The most exhaustive analysis is that of W92.
 His main results are that: (1) the cluster core, although
unresolved, appears to be blue; (2) the HB is bimodal, showing both a red
clump and an extended blue tail; (3) there are no radial trends in the
relative numbers of red horizontal branch (RHB), blue horizontal branch
(BHB) and red giant branch (RGB) stars for 48\arcsec~ $< r <$ 190\arcsec~;
(4) the RGB ``bump'' is at $V$ = 16.15 $\pm$ 0.03 mag; (5) the population ratio $R
= N({\rm HB})/N({\rm RGB})$ has a value 1.26 $\pm$ 0.10, which corresponds to a helium
abundance $Y$ = 0.23 $\pm$ 0.01 (computed by means of the R-method;
e.g. Renzini 1977); (6) there are six blue straggler (BS) stars and six
supra-RHB stars $[$15.7 mag $<$ $V$ $<$ 16.0 mag; 0.6 $< (B-V) <$ 0.8$]$ in the
region 120\arcsec~ $< r <$ 220\arcsec~, and there is evidence of
segregation only for the BS stars, so an origin for supra-RHB stars from
BS stars is not supported by W92 data; (7) no significant abundance spread
is found from the colour width of the main sequence (MS); and finally (8) an
age of 14 $\pm$ 1 Gyr results from the $\Delta$~($B$$-$$V$) method (Sarajedini
\& Demarque 1990; VandenBerg et al. 1990).

We have now obtained new {\sl large-field} CCD $V$, $I$ photometry for
NGC~1851. The new data set makes it possible to re-analyse the stellar
content of the cluster with a much richer sample and, for the first time,
allows a comprehensive study of its dynamical properties.  However, the
central regions of the cluster cannot be studied with this ground-based
material, due to the extreme crowding of the core. To overcome this
limitation, pre-repair {\it Hubble Space Telescope} ({\it HST}) images have
been retrieved from the archives and reduced in order to sample the central
stellar content, in particular the radial distribution of the HB stars. For
the sake of comparison, the photometric catalogue of W92 has been also used.

\begin{figure}[t]
\psfig{figure=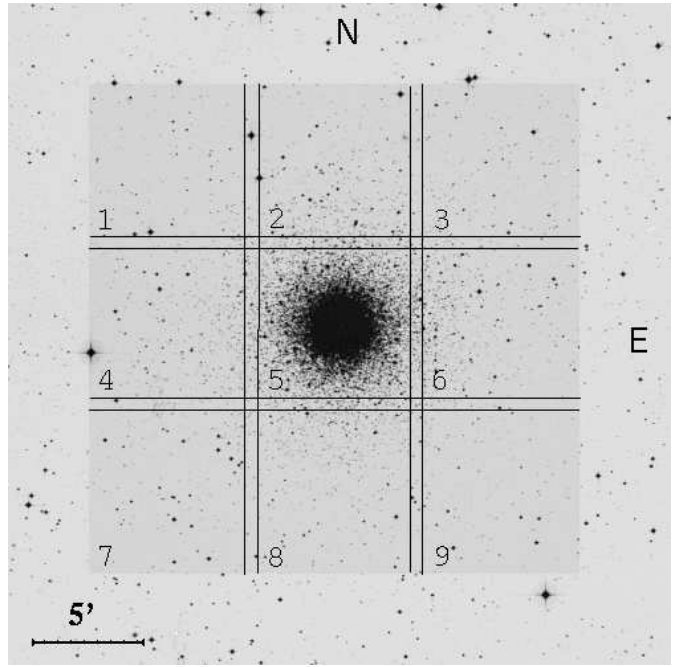,width=8.8cm}
\caption[]{ The observed NTT/EMMI fields, sketched over a POSS field.}
\label{field_map}
\end{figure}

\begin{small}
\begin{table}[t]
\caption[]{Log of NTT/EMMI observations.}
\label{observs}
\begin{tabular}{cccccccc}
\noalign{\smallskip}
\hline
\hline
\noalign{\smallskip}
  Nr.   &  Field & $t_{\rm exp}$(s) &  Filter & Date         & FWHM $[\arcsec]$
 \\
\noalign{\smallskip}
\hline
\noalign{\smallskip}
   1    &  6     & 50  &    $V$     &  1993 Feb 18  & 1.2 \\
   2    &  6     & 70  &    $I$     &  1993 Feb 18  & 1.2 \\
&&&&&\\
   3    &  5     & 45  &    $I$     &  1993 Feb 18  & 1.2 \\
   4    &  5     & 10  &    $I$     &  1993 Feb 18  & 1.4 \\
   5    &  5     & 10  &    $V$     &  1993 Feb 18  & 1.1 \\
   6    &  5     & 30  &    $V$     &  1993 Feb 18  & 1.3 \\
&&&&&\\
   7    &  2     & 30  &    $V$     &  1993 Feb 18  & 1.0 \\
   8    &  2     & 40  &    $I$     &  1993 Feb 18  & 1.2 \\
&&&&&\\
   9    &  3     & 60  &    $I$     &  1993 Feb 18  & 1.2 \\
  10    &  3     & 44  &    $V$     &  1993 Feb 18  & 1.1 \\
&&&&&\\
  11    &  1     & 45  &    $V$     &  1993 Feb 18  & 1.0 \\
  12    &  1     & 60  &    $I$     &  1993 Feb 18  & 1.2 \\
&&&&&\\
  13    &  4     & 60  &    $I$     &  1993 Feb 18  & 1.1 \\
  14    &  4     & 45  &    $V$     &  1993 Feb 18  & 1.1 \\
&&&&&\\
  15    &  7     & 45  &    $V $    &  1993 Feb 18  & 1.1 \\
  16    &  7     & 60  &    $I$     &  1993 Feb 18  & 1.2 \\
\noalign{\smallskip}
\hline
\noalign{\smallskip}
  17    &  8     & 70  &    $I$     &  1993 Feb 19  & 1.1 \\
  18    &  8     & 55  &    $V$     &  1993 Feb 19 & 1.1 \\
&&&&&\\
  19     &  9    & 55  &    $V$     &  1993 Feb 19 & 1.2 \\
  20     &  9    & 70  &    $I$     &  1993 Feb 19 & 1.2 \\
\noalign{\smallskip}
\hline
\noalign{\smallskip}
  21     &  back  & 120 &    $V$       & 1993 Dec 10  & 1.0 \\
  22     &  back  & 180 &    $I$       & 1993 Dec 10  & 1.1 \\
\noalign{\smallskip}
\hline
\end{tabular}
\end{table}
\end{small}

\section{The data set} \label{dataset}

\begin{figure}[t]
\psfig{figure=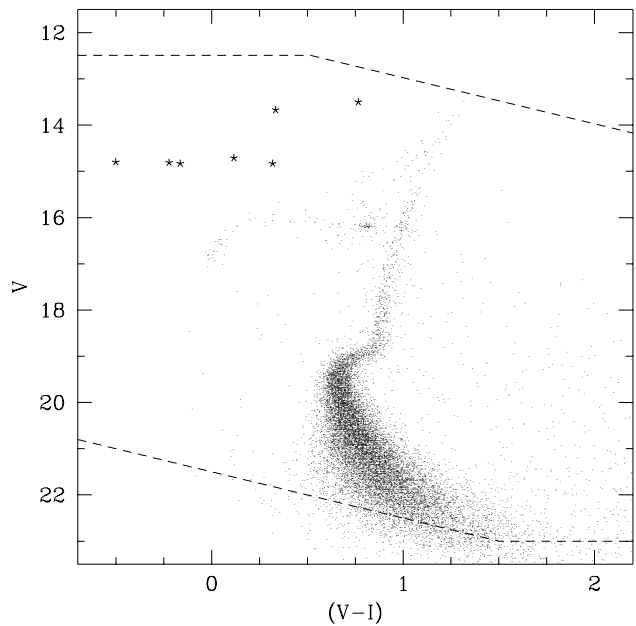,width=8.8cm}
\caption[]{
The CMD from the NTT/EMMI data: stars with a {\sc daophot}  $\chi < 1.5$
(Stetson 1987) are plotted.  The 50~\% completeness level and the
saturation line are shown (dashed lines), The bright-blue supra-HB
objects are marked with an asterisk.
}
\label{complete_cmd}
\end{figure}

The ground--based $V$, $I$ images have been obtained at ESO/La~Silla with
the 3.5~m New Technology Telescope (NTT) and the EMMI instrument at the
Nasmyth focus. The list of the NTT/EMMI exposures is given in Table
~\ref{observs}. Nine fields towards the cluster (1993 February) plus one
background field (1993 December) have been observed, in moderate to good
seeing conditions. The cluster fields cover a total area of 22 $\times$ 22
arcmin$^2$, which extends out to 1.3 tidal radii (cf. Sect.~\ref{lfmf}), and
are sketched in Fig.~\ref{field_map}.

\begin{table}[t]
\caption[]{List of {\it HST} images.}
\label{hstimages}
\begin{tabular}{rrclrl}
\noalign{\smallskip}
\hline\hline
\noalign{\smallskip}
\multicolumn{1}{c}{Nr.} &
\multicolumn{1}{c}{Date} &
\multicolumn{1}{c}{Instr.} &
\multicolumn{1}{c}{Filter} &
\multicolumn{1}{c}{$t_{\rm exp}$(s)} \\
\noalign{\smallskip}
\hline
\noalign{\smallskip}
        1  &      1991  Feb 12 & PC & F785LP &  160 \\
        2  &     1991  Feb 12  & PC & F555W  &  160 \\
\noalign{\smallskip}
\hline
\end{tabular}
\end{table}

In order to complete the radial coverage, the $HST$ archives have also been
scanned, and ten WFPC1 images of NGC~1851 were found.  
We reduced the two frames listed in Table~\ref{hstimages}, which have been
taken in bands close to the $V$ and $I$ bands observed from the ground.
These exposures cover the very centre of the cluster ($\sim$~30$''
\times$~30$''$).

A complete report of the reduction and calibration procedures is given
in the Appendices, to which the reader is referred for details. We
recall the main steps here.  All frames have been pre-processed in the
standard way, and the instrumental photometric catalogues have been
obtained with the {\sc daophot}+{\sc allstar} (Stetson 1987) codes. Subsequent
calibrations of the ground-based observations have been performed by
means of five Landolt (1992) standard stars for the February run, plus eight
standards for the December run. The {\it HST} CMD has been calibrated by
comparison to the NTT/EMMI CMD (see Appendix B for more
information). The overlap areas of the nine fields covering the
cluster have been used in order to estimate the photometric
zero-point accuracy (see Appendix A.2. for the details). Taking
into account both internal and external errors, we estimate a
zero--point error of 0.03 mag in the $V$ filter and 0.03 mag in the $I$
filter (1~sigma). 
The corresponding error in ($V$--$I$) is 0.04~mag. 
These figures apply also to
the zero--points of the {\it HST} CMD.

Both completeness and the photometric errors have been evaluated with
artificial star tests: the stars have been created with colours and
luminosities representative of the MS--RGB sequence. The results of
these tests are summarized in Table 11, which shows that
the internal error keeps to within 0.1 mag until $V$ $\simeq$ 21 mag. A
discussion of the completeness is deferred to Sect.~\ref{lfmf} in the
context of the analysis of the luminosity and mass functions.

\section{The colour--magnitude diagram} \label{thecmd}

\begin{figure}[t]
\psfig{figure=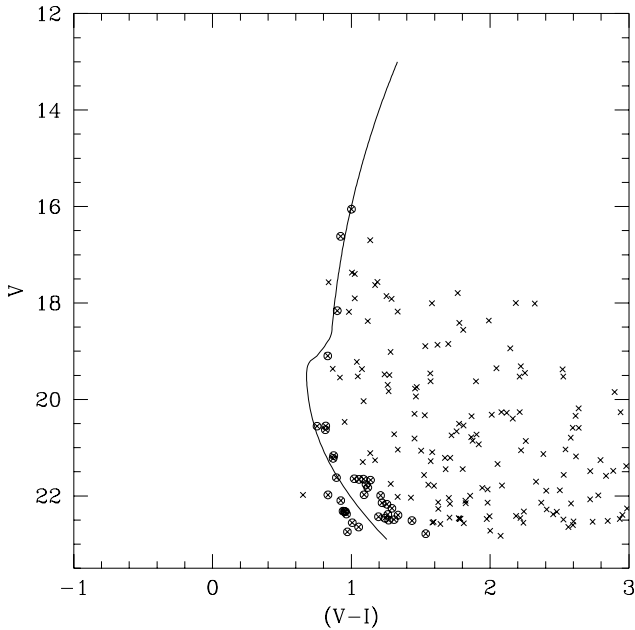,width=8.8cm}
\caption[]{The CMD of the background field. This  is located at
30\arcmin~ east of the cluster centre
and covers 9.6\arcmin~ $\times$ 7.4\arcmin~. The solid line shows the
fiducial points of NGC~1851, and stars within 2$\sigma$ of the line are
further marked with open circles.}
\label{background_cmd}
\end{figure}

The calibrated CMD of NGC~1851 from the NTT data is presented in
Fig.~\ref{complete_cmd}, while Fig.~\ref{background_cmd} shows the CMD
of the background field. In order to better highlight the CMD features,
only stars with {\sc allstar} parameter $\chi$ $\leq$~1.5 are plotted in
Fig.~\ref{complete_cmd} ($\simeq 20\,000$ stars). Here, the dashed
lines mark the 50~\% completeness level and the saturation limit:
within these limits, the photometry extends from $\sim$~1 mag below
the RGB tip to $\sim$ 4 mag below the turn-off (TO).
The richness of the sample allows the identification of the well populated
main sequence, subgiant branch and giant branch, as well as the less
represented horizontal branch, blue straggler sequence and asymptotic giant
branch (actually, two AGBs can be seen in Fig.~\ref{complete_cmd}). In the
following sections, we will carry out a detailed comparison of each sequence
with current stellar models, and we first need to establish the cluster's
fundamental parameters: reddening, distance and metallicity.

\subsection{Absolute calibration of the CMD} \label{abscalibration}

The most reliable estimate of the cluster reddening has been obtained by
S81, using Str\"omgren photometry of the foreground stars. He
determined $E_{B-V}$ = 0.02$\pm$0.01. A low value of the reddening is
also to be expected considering the relatively great angular distance of NGC~1851
from the Galactic plane ($b = -35^\circ$): indeed,
on the basis of the maps of Burstein \& Heiles (1982), we would expect $E_{
B-V}$ = $0.015 \pm 0.02$.
Therefore, we will adopt $E_{B-V}$ = 0.02; the extinction curve from
Savage \& Mathis (1979) then gives $A_V$ = 0.06, $E_{V-I}$ = 0.03 and $A_I$
= 0.03.

No direct estimates of the metallicity of NGC~1851 from high-dispersion
spectroscopy exist. Current determinations are therefore based on a variety
of spectroscopic and photometric metallicity indices, obtained both from
integrated-light and single-star measurements. A brief account of recent
determinations is now given.

Zinn (1980) measured the photometric metallicity index $Q_{39}$ from
integrated spectra. By placing his sample of clusters on the high-dispersion
metallicity scale of Cohen (1978a,b; 1979a,b) he obtained a value [Fe/H]$ =
-1.33$ from the NGC~1851 $Q_{39}$ value.

Frogel et al. (1983) obtained integrated infrared  $(J-K)_0$ and
$(V-K)_0$ colours of individual RGB stars of the cluster, which they used as
metallicity indices calibrated on the new high--dispersion
measurements of Cohen (1982, 1983). They estimated a metallicity [Fe/H]$ =
-1.26$.

Zinn \& West (1984, ZW84) used the $Q_{39}$ value of Zinn (1980) and Cohen's
new metallicity scale to re-evaluate NGC~1851's metallicity, which they
found to be [Fe/H]$ = -1.34 \pm 0.09$.

Armandroff \& Zinn (1988) measured the equivalent widths of the infrared
Ca~{\sc ii} triplet on integrated spectra of the cluster. They used the
scale of Zinn \& West (1984) to rank their sample of clusters and obtained a
value [Fe/H]$ = -1.16$.

Da Costa \& Armandroff (1995, DA95) repeated the measurement of the infrared
equivalent widths, in this case using individual cluster RGB stars.
These equivalent widths have recently been used by Rutledge et al. (1997) to
obtain a new estimate of the cluster's metallicity (DA95 used NGC~1851 as a
calibration cluster). Depending on the method followed, Rutledge et al.
find [Fe/H]$ = -1.33 \pm 0.09$ or $-1.23 \pm 0.11$ if using the ZW84 scale,
or [Fe/H]$ = -1.03 \pm 0.06$ on the Carretta \& Gratton (1997, CG97) scale.

Taking the previous determinations as independent measurements of the
NGC~1851 metallicity, we find a mean value [Fe/H]$= -1.28 \pm 0.07$,
obtained excluding the value on the CG97 scale. Using this datum we
find [Fe/H]$= -1.24 \pm 0.11$, which is entirely compatible with the
previous value. On the other hand, until new consensus is reached on the
CG97 scale, and for consistency with the other determinations the slightly
higher value will be adopted in the following. We furthermore stress that our
conclusions are by no means affected by the choice of either value.

\begin{table}[t]
\caption[]{Summary of distance determinations.}
\label{hb_level}
\begin{tabular}{lllll}
\noalign{\smallskip}
\hline\hline
\noalign{\smallskip}
Ref. & $V_{\rm HB}$ & $A_{V}$ & $M_{V}^{\rm HB}$ & $(m-M)_0$ \\
\noalign{\smallskip}
\hline
\noalign{\smallskip}
1 & 15.65           & 0.72            & 0.5 & 14.43            \\
2 & 16.05 $\pm$ 0.1 & 0.06 $\pm$ 0.04 & 0.6 & 15.39 $\pm$ 0.23 \\
3 & 16.05 $\pm$ 0.2 & 0.06            & 0.6 & 15.39 $\pm$ 0.20 \\
4 & 15.80           & 0.42            & 0.5 & 14.88 $\pm$ 0.22 \\
5 & 16.10           & 0.30            & 0.6 & 15.20            \\
6 & 16.10           & \multicolumn{2}{c}{isochrone fit} & 15.45 \\
\noalign{\smallskip}
\hline
 1 & \multicolumn{4}{l}{ Alcaino (1969)             } \\
 2 & \multicolumn{4}{l}{ Alcaino (1971)             } \\
 3 & \multicolumn{4}{l}{ Racine (1973)              } \\
 4 & \multicolumn{4}{l}{ Burstein \& McDonald (1975)} \\
 5 & \multicolumn{4}{l}{ Alcaino (1976)             } \\
 6 & \multicolumn{4}{l}{ Harris (1976)              } \\
\noalign{\smallskip}
\hline
\end{tabular}
\end{table}

\begin{table}[t]
\caption[]{Data for the calculation of distance from HB luminosity:
here $a$ and $b$ refer to the
relation $M_{V}^{\rm HB}$ = $a$~[Fe/H] + $b$. We assume [Fe/H] = --1.28,
$A_{V}$ = 0.06 and $V_{\rm HB}$ = 16.2 mag. An asterisk marks the data
corrected as explained in the text.}
\label{mod_distanza}
\begin{tabular}{ccccl}
\noalign{\smallskip}
\hline\hline
\noalign{\smallskip}
  Ref.   &     $a$   &     $b$  &  $M_{ V}^{\rm HB}$ & $(m-M)_0$  \\
\noalign{\smallskip}
\hline
\noalign{\smallskip}
   1  &   0.19  &   0.91  &   0.67 &   15.47 \\
   2  &   0.15  &   0.44  &   0.25 &   15.89 \\
   3  &   0.39  &   1.17  &   0.67 &   15.47 \\
   4  &   0.22  &   1.02  &   0.74 &   15.40 (*)\\
   5  &   0.20  &   1.21  &   0.95 &   15.19 \\
   6  &   0.20  &   0.93  &   0.67 &   15.47 (*)\\
   7  &   0.51  &   1.58  &   0.93 &   15.21 \\
\noalign{\smallskip}
\hline
\noalign{\smallskip}
  1	& \multicolumn{4}{l}{Rood \& Crocker (1989): theoretical} \\
  2	& \multicolumn{4}{l}{Fusi Pecci et al. (1990): RGB bump} \\
  3	& \multicolumn{4}{l}{Sandage \& Cacciari (1990): period-shift} \\
  4	& \multicolumn{4}{l}{Lee et al. (1990): theoretical} \\
  5	& \multicolumn{4}{l}{Carney et al. (1992): Baade--Wesselink} \\
  6	& \multicolumn{4}{l}{Walker (1992b): LMC Cepheids + RR Lyr} \\
  7	& \multicolumn{4}{l}{Rees (1993): statistical parallaxes} \\
\noalign{\smallskip}
\hline
\end{tabular}
\end{table}

The previous distance determinations for NGC 1851 are less than
satisfactory. Indeed, Table~\ref{hb_level} shows the distance moduli
available in the literature: they range from $(m-M)_0=14.43$ (Alcaino
1969) to $(m-M)_0=15.45$ (Alcaino et al. 1990), with 
$< (m-M)_0 > =15.12 \pm 0.36$.

Almost all the above determinations of the distance
(cf. Table~\ref{hb_level}) are based on the luminosity of the HB
compared to its absolute luminosity, where the latter is assumed not
to vary with metallicity. Here we re-evaluate the distance allowing
the absolute visual magnitude of the ZAHB, to
vary with the metal content according to the law

\smallskip
\noindent $M_{V}^{\rm ZAHB}$ = $a$~[Fe/H] + $b$,
\smallskip

\noindent where the values of the parameters $a$ and $b$ are still
controversial. Table~\ref{mod_distanza} shows some recent
determinations of these parameters.  Accordingly, the distance modulus
has been computed adopting, from the previous discussion, [Fe/H] =
--1.28 and $A_{V}$ = 0.06.

The peak in the $V$ distribution of the HB stars is at $V$ = 16.2 $\pm$
0.20 mag (see Fig.~\ref{complete_cmd}).
Since the number of stars in a region of the CMD is
proportional to the lifetime of the corresponding evolutionary phase,
the stars belonging to this peak must be identified with the
longest-lived phase of an HB star, which corresponds to a region
close to the ZAHB.  We note that the relation reported
in Table~\ref{mod_distanza} should strictly hold for ZAHB stars. In
those cases in which the authors give a relation for RR~Lyr stars,
the coefficients were corrected following Carney et al. (1992), who
find:

\smallskip
\noindent $M_{V}^{\rm RR}$ = $M_{V}^{\rm ZAHB}$ -- 0.05~[Fe/H] -- 0.20
\smallskip

\noindent
(see also Catelan 1992). Corrected data are marked with an asterisk in
Table~\ref{mod_distanza}. A mean of the listed moduli gives $(m-M)_0$ =
15.44 $\pm$ 0.20, which also results from the isochrone fitting (see
below).  This is the distance modulus we will adopt in the following.

\subsection{Main sequence and red giant branch}

\begin{table}[t]
\caption[]{Fiducial points and widths along the MS--SGB--RGB sequence.}
\begin{tabular}{cccccc}
\noalign{\smallskip}
\hline\hline
\noalign{\smallskip}
\multicolumn{2}{c}{$V$ range} &  $<V$--$I>$ & $\sigma$ & $(V-I)_{\rm max}$ & Bin \\
\noalign{\smallskip}
\hline
\noalign{\smallskip}
	15.50 &	15.75 &	1.03  & 0.03 &	1.02  & 0.02 \\
	15.75 &	16.00 &	1.01  & 0.02 &	1.02  & 0.02 \\
	16.00 &	16.25 &	0.99  & 0.02 &	1.00  & 0.02 \\
	16.25 &	16.50 &	0.98  & 0.03 &	0.98  & 0.02 \\
	16.50 &	16.75 &	0.94  & 0.02 &	0.94  & 0.02 \\
\noalign{\smallskip}
	16.75 &	17.00 &	0.95  & 0.02 &	0.94  & 0.02 \\
	17.00 &	17.25 &	0.92  & 0.02 &	0.92  & 0.02 \\
	17.25 & 17.50 &	0.90  & 0.03 &	0.90  & 0.02 \\
	17.50 &	17.75 &	0.90  & 0.03 &	0.90  & 0.02 \\
	17.75 &	18.00 &	0.88  & 0.03 &	0.88  & 0.02 \\
\noalign{\smallskip}
	18.00 &	18.25 &	0.89  & 0.03 &	0.90  & 0.02 \\
	18.25 &	18.50 &	0.86  & 0.03 &	0.86  & 0.02 \\
	18.50 &	18.75 &	0.85  & 0.03 &	0.84  & 0.02 \\
	19.00 &	19.25 &	0.70  & 0.08 &	0.67  & 0.03 \\
	19.25 &	19.50 &	0.68  & 0.07 &	0.68  & 0.02 \\
\noalign{\smallskip}
	19.50 &	19.75 &	0.68  & 0.07 &	0.67  & 0.03 \\
	19.75 &	20.00 &	0.68  & 0.06 &	0.66  & 0.02 \\
	20.00 &	20.25 &	0.71  & 0.07 &	0.70  & 0.03 \\
	20.25 &	20.50 &	0.72  & 0.08 &	0.72  & 0.04 \\
	20.50 &	20.75 &	0.76  & 0.09 &	0.76  & 0.04 \\
\noalign{\smallskip}
	20.75 &	21.00 &	0.77  & 0.10 &	0.75  & 0.05 \\
	21.00 &	21.25 &	0.83  & 0.13 &	0.85  & 0.05 \\
	21.25 &	21.50 &	0.87  & 0.15 &	0.88  & 0.06 \\
	21.50 &	21.75 &	0.90  & 0.15 &	0.88  & 0.06 \\
	21.75 &	22.00 &	0.95  & 0.18 &	0.94  & 0.07 \\
\noalign{\smallskip}
	22.00 &	22.25 &	1.02  & 0.18 &	1.04  & 0.08 \\
	22.25 &	22.50 &	1.04  & 0.18 &	1.04  & 0.08 \\
	22.50 &	22.75 &	1.15  & 0.21 &	1.12  & 0.08 \\
	22.75 &	23.00 &	1.28  & 0.26 &	1.33  & 0.09 \\
\noalign{\smallskip}
\hline
\noalign{\smallskip}
\multicolumn{6}{c}{Turn-off} \\
\noalign{\smallskip}
\multicolumn{2}{c}{$V$--$I$ range} &  $<$$V$$>$ & $\sigma$ & $V_{\rm max}$ & Bin \\
\noalign{\smallskip}
\hline
\noalign{\smallskip}
	 0.65 &	 0.70 &	19.26  & 0.17 &	19.20 & 0.15 \\
	 0.70 &	 0.75 &	19.13  & 0.15 &	19.20 & 0.15 \\
	 0.75 &	 0.80 &	19.03  & 0.13 &	19.00 & 0.10 \\
	 0.80 &	 0.85 &	18.82  & 0.11 &	18.80 & 0.10 \\
	 0.85 &	 0.90 &	18.62  & 0.23 &	18.60 & 0.20 \\
\noalign{\smallskip}
\hline
\noalign{\smallskip}
\end{tabular}

\label{fid_points}
\end{table}

In order to establish our fiducial lines and to compare the observed
widths with the photometric errors, a careful analysis of the
colour distribution along the MS--SGB--RGB sequence has been made. We
assumed a Gaussian colour distribution and determined the mean colour
and dispersion, $\sigma$, in each magnitude interval: these data are
grouped in Table \ref{fid_points}, where the mode of the distribution
and the bin width (1 $\sigma$) are also listed.

Once the observed widths of Table \ref{fid_points} are compared with
the $\sigma$ associated with the photometric errors
(Table 11), it can be seen that the former are compatible
with a null intrinsic colour dispersion along the MS--SGB--RGB sequence
that translates into a null dispersion in metallicity (see
e.g. Renzini \& Fusi~Pecci 1988).

\subsection{The RGB bump}

\begin{figure}[t]
\psfig{figure=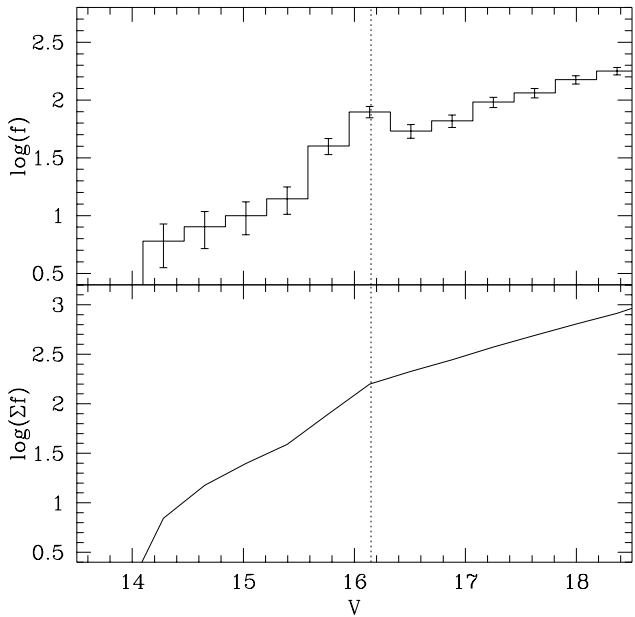,width=8.8cm}
\caption[]{
Luminosity function (LF) for the evolved stars in NGC~1851: 
the differential LF is shown in the
upper panel; the cumulative LF is plotted in the lower panel. The bump in the
LF is clearly seen in both the differential and cumulative LF
at $V$ = 16.15 $\pm 0.15$ mag.}
\label{v_bump}
\end{figure}

\begin{figure}[t]
\psfig{figure=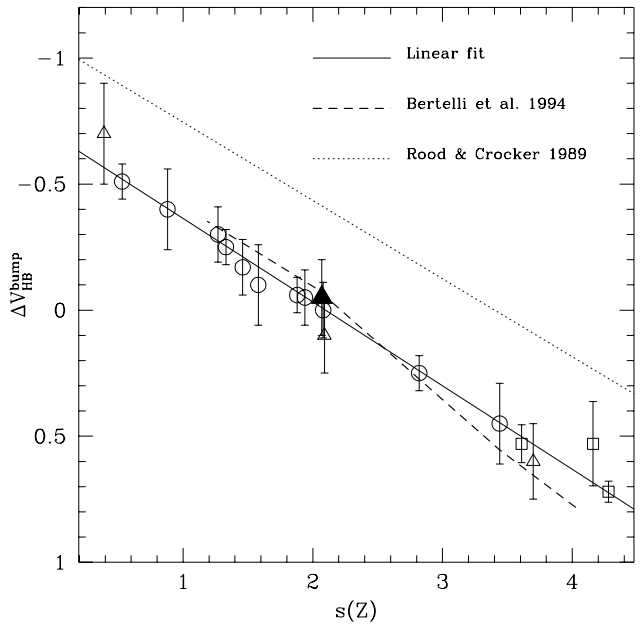,width=8.8cm}
\caption[]{
The quantity $\Delta V_{\rm bump}^{\rm HB}$ as a function of the parameter
$s(Z)$ = sinh$^{-1}$ ($Z_4$/2.5) (see text); a linear fit to the data
yields $\Delta V_{\rm bump}^{\rm HB}$ = 0.33 $s(Z) - 0.70$ (0.06 rms).  The filled
triangle represents NGC~1851 and the open triangles represent M~80,
M~4 and NGC~6352 (Saviane \& Piotto in prep.); the open circles are the
original sample of Fusi Pecci et al. (1990) and the open squares are 3
other clusters from Sarajedini \& Norris (1994).  Also represented are
the linear fit (solid line) and the theoretical relations from
Bertelli et al. (1994, dashed line) and Rood \& Crocker (1989, dotted
line). }
\label{bumps}
\end{figure}

The RGB bump was first predicted by Iben (1968) in the evolutionary phase of
H-burning in a shell: the bump is located on the RGB at a luminosity which
depends on the metallicity and on the age of the cluster, and generally
speaking it is not a prominent feature. For example, Iben (1968) predicted
that the ratio of lifetimes between the bump and the HB phases is $t_{\rm
bump} / t_{\rm HB} = 0.025$ and that the ratio between the bump and RGB
phases is $t_{\rm bump} / t_{\rm RGB} = 0.035$ for $Y = 0.35$ and $Z = 2
\cdot 10^{-4}$; therefore, for every 100 RGB stars just $\sim 4$ will be
found in the bump (slightly higher figures are obtained for lower $Y$
values). Moreover, in metal-poor clusters the bump is located in the bright
part of the RGB, where the statistics are poor and it can be lost in the
count noise.  For this reason, it has not been detected until recently (King
et al. 1985). Afterwards, it has been studied on a sample of GCs by
Fusi~Pecci et al. (1990, FP90), and the sample has been further extended by
Sarajedini
\& Norris (1994).  As shown in Fig.~\ref{v_bump}, the bump in the RGB of
NGC~1851 is located at $V$ = 16.15 $\pm$ 0.15 mag.  Assuming $V_{\rm HB}$ = 16.20
$\pm$ 0.03 mag (cf. Sect.~\ref{abscalibration}) we have obtained $\Delta
V_{\rm bump}^{\rm HB} = -0.05 \pm 0.15$.

FP90 provide a relative calibration for $\Delta
V_{\rm bump}^{\rm HB}$ which corresponds to the difference between the
absolute magnitude of the bump ($M_{V}^{\rm bump}$) with respect to the
luminosity of the HB ($M_V^{\rm HB}$), for a sample of 11~globular
clusters.  The value determined for NGC~1851 is in good agreement with
the above calibration. In fact, we have extended this calibration with the
inclusion of seven additional clusters from Sarajedini \& Norris (1994, three
clusters), Saviane \& Piotto (in prep., three clusters) and NGC~1851 (this
paper).  We find that the best fit is given by the following relation:

\[
\Delta V_{\rm bump}^{\rm HB} = 0.33 \sinh^{-1} \left( \frac{Z_4}{2.5} \right) - 0.70
\,\,\, (0.06 \,\, {\rm rms}).
\]

The parameter sinh$^{-1}$~($Z_4$/2.5) is chosen in order to linearize the
dependence on the metallicity; $Z_4$ is the metallicity in units of 0.0004.
Since the dependence of the luminosity of the HB on [Fe/H] accounts only
for a variation of 0.3--0.4~mag in $\Delta V_{\rm bump}^{\rm HB}$ in the
metallicity interval covered by our data, Fig.~\ref{bumps} shows that the
bump goes fainter when higher [Fe/H] values are considered. It is also evident
that NGC~1851 (filled triangle) follows the general trend.

As far as the theoretical interpretation is concerned, we first
considered the canonical models by Rood \& Crocker (1989) whose
relation is given in FP90 assuming a constant age (15~Gyr) and helium
abundance ($Y = 0.23$). In Fig.~\ref{bumps} this relation is shown as
a dotted line, and it can be seen that its slope is similar to the
observational one, but there is a clear luminosity difference, in the
sense that the theoretical values are brighter than the empirical ones
($\sim 0.4$ mag for the lower metallicities and $\sim 0.5$~mag for the
higher ones).  Some attempts to solve this problem are examined below.

On the basis of spectroscopic observations of giant stars in globular
clusters (e.g. Gratton \& Ortolani 1989), which show that in these stars
the $\alpha$ elements are enhanced with respect to the solar values,
Salaris et al. (1993) presented Population~II isochrones computed with such
non--standard chemical compositions.  One of the main properties of
$\alpha$-enhanced isochrones is that they are well reproduced by
solar-scaled isochrones.  If we take an isochrone of metallicity $Z_0$
whose $\alpha$-elements are enhanced by a factor $f$, it can be
represented by a solar-scaled isochrone of a higher metallicity $Z = Z_0
\, (0.638 \, f + 0.362)$.  A solar-scaled isochrone is one with
solar metallic partition and scaled in global metallicity from $Z_0$ (that
of the alpha-enhanced metallic partition) to $Z$, according to the above
formula.  Straniero et al. (1992) suggest that these kinds of models could
solve the RGB bump discrepancy, since a higher metallicity would make the
bump fainter and therefore $\Delta V_{\rm HB}^{\rm bump}$ fainter as well.

In order to quantify this effect, we can use the observed relation
$\Delta V_{\rm HB}^{\rm bump}$ vs. [Fe/H]. Even if it is not exactly
linear, a good approximation is $\Delta V_{\rm HB}^{\rm bump} =
0.66$~[Fe/H]$ + 0.87$. The relation between $Z$ and $Z_0$ can be
transformed into [Fe/H]~$\simeq$~[Fe/H]$_0 - \log (0.638\,f +0.362)$.
All current estimates give an upper limit of $f
\simeq 4$ (see the compilation in Salaris et al. 1993), and even
adopting this extreme value we find that the theoretical $\Delta
V_{\rm HB}^{\rm bump}$ can be made only $\simeq$0.3 mag fainter.

Another approach is to consider that a certain amount of convective
overshoot occurs at the bottom of the envelope of low-mass stars while
climbing the RGB (King et al. 1985; Alongi et al. 1991).  In this set
of models the envelope goes deeper into the innermost nuclear
processed regions of the star, hence the first dredge-up is
anticipated. As a consequence, the bump is moved $\sim$ 0.4 mag
fainter. The isochrone library by Bertelli et al. (1994) assembles
stellar models adopting this prescription and the dashed line in
Fig.~\ref{bumps} shows the resulting fairly good agreement with the
observed luminosities in the metal range covered. As for the relation
labelled ``Rood \& Crocker 1989'' in Fig.~\ref{bumps}, also the relation
labelled ``Bertelli et al. 1994'' has been derived assuming a constant
age of 15~Gyr. The helium content is not constant but
follows the ratio $\Delta Y/\Delta Z = 2.5$. This means higher helium
content for metal-rich clusters.

We note a possible residual offset at the high- and low-metallicity ends
between the data and the Bertelli et al. (1994) relation, specifically the
theoretical relation seems to predict lower values for $\Delta V_{\rm
HB}^{\rm bump}$ at $s(Z) < 2.2$ and higher values at $s(Z) > 2.2$.  The
discrepancy depends either on the HB or on the bump luminosities. In the
first case, Fig.~\ref{bumps} shows that the HB luminosity is too high for
metal-rich clusters, and too low for metal-poor ones. Conversely, in the
second case the bump occurs at luminosities that are too low for the metal-rich 
clusters and too high for the metal-poor ones.

The HB luminosity is
mostly controlled, at fixed metallicity, by the helium content.  Higher $Y$
means brighter HB luminosities. A different (lower)  $\Delta
Y/\Delta Z$ ratio would act in the right direction, lowering the HB luminosity at
high $Z$ values and enhancing it at low $Z$ values.

It should also be remembered that the luminosity of the RGB bump critically depends,
at fixed metallicity, on the age. Specifically, lower ages imply
higher RGB stellar masses which in turn possess higher luminosities.
As a consequence, an interesting way of removing the cited offsets would
be to assume that high-metallicity clusters are younger than low-metallicity 
ones. To quantify the required corrections, we computed
values for the age differences at the extremes of the metallicity
range covered by the models. We found that for [Fe/H]~=~--0.4 the
clusters should have an age $\simeq 3.3$~Gyr lower and for
[Fe/H]~=~--1.74 they should have an age $\simeq 1.6$~Gyr higher than
the age of metal-intermediate clusters. The present comparison would
therefore suggest an age-metallicity relation among globular
clusters, with an age difference of $\simeq 5$~Gyr between metal-poor
and metal-rich objects. However, we caution the reader that in any
case the uncorrected relation is fairly compatible with the
observations given the associated errors, and the quoted age
difference could be considered as an upper limit.  In summary, we
argue that both $\alpha$-enhanced chemical composition and envelope
overshoot could actually exist in real GC stars. The former is
observational evidence, but even considering this effect in
theoretical models, the discrepancy seems to be not completely
removed, thus some degree of overshoot must be invoked. Therefore, the
suggested age-metallicity relation should be more carefully probed
with improved models including both effects.

\subsection{AGB and supra horizontal branch stars} \label{sect_shb}

\begin{figure}

\psfig{figure=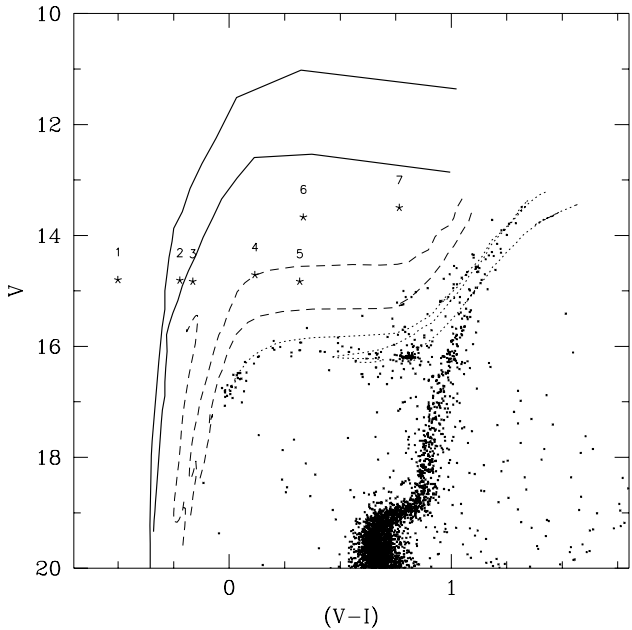,width=8.8cm}

\caption[]{
Enlarged portion of the CMD showing in better detail the horizontal branch
region. Stars brighter than the HB and bluer than the AGB are marked with
an asterisk and numbered (supra-HB stars). The solid lines are the
``planetary nebula'' part of two isochrones, computed for an age of 15~Gyr
and two different mass losses along the RGB (see text). The other lines
correspond to the tracks of stars evolving off-HB, for different initial
HB masses and compositions ($Z$=0.001 for the dotted lines and $Z$=0.0004 for
the dashed lines). Going from bluer to redder colours the initial HB masses
are 0.50, 0.52 and 0.55~$M_\odot$ for the lower metal content, and 0.60,
0.65 and 0.70~$M_\odot$ for the higher metallicity.  The $V$ magnitude of the
plateau of the post-HB tracks is roughly the same for the two adopted
metallicities.  The majority of supra-HB stars can be identified with some
post-HB evolutionary phase.
}
\label{supra_teo}
\end{figure}

A number of stars are found in the CMD area brighter than the HB and bluer
than the AGB (the so-called ``supra-HB'' stars), and in
Fig.~\ref{cm_supra_hst} (upper panel) a dotted line marks the limit of what
we considered supra-HB candidates, namely stars 1 mag brighter than
the HB and bluer than the AGB.  We have found $\sim$ 50 stars in this
region.  They have been inspected on the original frames and catalogued
according to (a) the {\sc allstar} photometric parameters, (b) the product and
the ratio of the FWHM and (c) the value of the brightest pixels.  17 stars
are found to have some saturated pixels in the I filter, and 16 more are
blends (the product of the FWHMs is more than 2 times the mean one, or the
sum of the $\chi$ parameters is greater than 10). Of the remaining group, 7
stars have a sum of the {\sc daophot} $\chi$ in the two filters less than 4, and
therefore can be considered objects with good photometry, while the
remaining 11 stars, with 4 $<$ $\chi$ $<$ 10, are probable blends and hence
discarded. The seven objects which passed this selection are marked with an
asterisk in Fig.~\ref{complete_cmd}; they present normal star-like shape,
do not have saturated pixels and are not in blends.
 
From Fig.~\ref{background_cmd} we see that hardly any field stars are
expected in this CMD area, and indeed the Bahcall \& Soneira model
(Bahcall 1986; adapted to the ($V$--$I$) colour as in Saviane et al. 1996)
predicts at most 1 star for a colour range ($V$--$I$) $<$ 0.5, within an
area of 484 arcmin$^2$ in the direction $l = 244^\circ$, $b = - 35^\circ0$;
thus, most of the supra-HB objects are likely stars of the cluster.
In order to identify their evolutionary status, we plotted in
Fig.~\ref{supra_teo} an enlargement of the HB region where supra-HB
stars have been numbered, and a number of isochrones and
tracks  superimposed.  The dotted lines represent the post-HB evolutionary tracks for
initial HB masses 0.60, 0.65 and 0.70~$M_\odot$ taken from the Padua
library with $Z$=0.001. The theoretical plane has been translated into
the observational one through the library of stellar spectra by Kurucz
(1992).  These tracks are able to explain the evolution
of HB and later phases for stars on the blue tail (0.6~$M_\odot$
tracks), intermediate HB (0.65~$M_\odot$) and red HB (0.7~$M_\odot$),
and to partially reproduce the observed dispersion in colour along the
AGB. This range of masses is larger than what one would expect from
normal mass-loss rates and their dispersion along the RGB. A
deeper analysis of this subject is given in Sect.~\ref{thehb}.

Since the library we are using does not contain models for HB masses lower
than 0.6~$M_\odot$ and $Z$=0.001, the evolution of smaller HB masses can be
explored by using the $Z$=0.0004 models and the bolometric corrections
pertaining to them, even if this value of $Z$ is less than the assumed
cluster metallicity. In fact, comparing the 0.6~$M_\odot$ tracks for the
two metallicities we verified that their differences in effective
temperatures and luminosities are not dramatic. On the other hand, when
applying the bolometric corrections, we find that the 0.06~dex difference
in the ZAHB $\log T_{\rm eff}$ implies that the $Z$=0.001 model starts $\sim
0.4$~$V$~mag brighter than the Z=0.0004 model, while {\sl the absolute {\rm V}
magnitude of the post-HB ``plateau'' is roughly unchanged}. 
Therefore, the
reader should not give too much importance to the fact that in
Fig.~\ref{supra_teo} the lower--mass evolutionary tracks start in an HB
region almost devoid of stars. There are in fact two  effects
that should be taken into account. First, a choice of the correct
metallicity would make the HB starting point brighter; secondly, theoretical
bolometric corrections could be slightly overestimated at these
temperatures.

These lower-metallicity tracks (dashed lines in Fig.~\ref{supra_teo})
are for initial HB masses 0.50, 0.52 and 0.55~$M_\odot$, going from the
bluer to the redder one.  We first note that this set of $M <
0.6\,M_\odot$ tracks is able to reproduce the bluer AGB, which is seen as a
separate branch running parallel to the higher-mass AGB, already discussed
(see Fig.~\ref{supra_teo}).  It is also evident that the location of stars
4 and 5 is compatible with the 0.52~$M_\odot$ track, in the post-HB phase
approaching the AGB.  Furthermore, we notice that (a) the post-HB
evolution of the 0.50~$M_\odot$ model is quite different from that of the
0.52~$M_\odot$ one, and (b) a decrement in mass as small as 0.03~$M_\odot$
between the two redder models increases their post-HB luminosity by $\sim
1$~mag. The 0.50~$M_\odot$ model evolves directly from the HB to the
planetary nebula (PN) regime, skipping the AGB phase (AGB-manqu\'e;
e.g. Fagotto et al. 1994 and references therein).  In the cited mass
interval we therefore expect that, for decreasing masses, a brighter and
brighter post-HB phase will be reached, until the star envelope is so much
reduced that the AGB cannot be developed.  From this point on, the
resulting AGB-manqu\'e phase will attain lower and lower luminosities.  On
the basis of these arguments, it is possible that a model with initial HB
mass $0.50\ M_\odot < M < 0.52\ M_\odot$ be able to develop the AGB after passing a
post-HB phase as luminous as the level of stars 6 and 7.

Instead, stars 2 and 3 can be identified as objects in the PN phase
either evolving from the classical post-AGB or in the final
AGB-manqu\'e phase. The first case is shown by the solid lines in
Fig.~\ref{supra_teo}, which are the PN portion of two isochrones,
computed for an age of 15~Gyr and a RGB mass-loss efficiency in the
Reimers formalism $\eta = 0.25$ (the brighter one) and $\eta = 0.43$
(the fainter one). These efficiencies have been chosen after the
analysis in Sect.~\ref{thehb}, and represent the reasonable boundaries
which are required in order to explain the blue and red part of the
HB.  The corresponding core masses, $M_{\rm c}$, for the two PNe are
$M_{\rm c} = 0.581 \, M_\odot$ and $M_{\rm c} = 0.537 \, M_\odot$.
The second case would be given by a track similar to the
0.50~$M_\odot$, $Z$=0.0004 model with a slightly higher initial HB mass.

Star 1 seems not to be compatible with either of the above
predictions, since its ($V$--$I$) colour ($\simeq -0.5$) is bluer than that
of the maximum blue excursion of PNe, which are the bluest objects
expected for this cluster.  There is a remote possibility that star 1
is a foreground object. Its colour is compatible with extreme values
that are observed for field stars. For example, the catalogue of Landolt
(1992), contains two objects (SA~107~215 and PG~0231~051), whose ($V$--$I$)
colours ( --0.511 and --0.534, respectively) are of the same order of
the one of star 1. The presence of 1 star in this CMD area is in
agreement with the field-star contamination previously evaluated.

A further test giving clues on the nature of the supra-HB stars,
would be to compare the relative counts of HB and post-HB objects to
those observed in globular clusters. However, in our case this test
is not significant.  In fact, we estimated the ratio $N$(post-HB)/
$N({\rm HB})$ to be $0.18 \pm 0.04$, which is compatible with the
ratio $R_2 = N({\rm AGB})/N({\rm HB}) = 0.15 \pm 0.05$ (Buzzoni et al. 1983)
determined on a sample of GCs. Excluding the seven supra-HB stars,
 our ratio is only slightly lowered and hence
still compatible with the observed $R_2$.

Finally, if supra-HB stars are the descendants of HB stars we also expect
them to follow the same radial distribution. Unfortunately, there are
too few supra-HB stars in our sample to do any statistical test,
while they are saturated in the {\it HST}/WFPC1 images.

\subsection{RR Lyrae stars} \label{rr_section}

\begin{table}[t]
\begin{scriptsize}
\begin{flushleft}
\caption[]{Positions and photometry of RR Lyrae stars. The variable type
($ab$ or $c$) is taken from the references numbered as follows: (1) Liller
(1975), (2) Wehlau et al. (1978) and (3) Wehlau et al. (1982). Other
characteristics taken from these references are labelled ``n'', ``bl.'', ``lP'',
``fld'' indicating, respectively, non-variable, blend, long period and field
star. Star 24 has a possible period of 183 days and is labelled ``183''.}
\label{rr_lyrae}
\begin{tabular}{rrrrrrll}
\hline \hline \noalign{\smallskip}
\multicolumn{1}{c}{$N$} &
\multicolumn{1}{c}{$x''$} &
\multicolumn{1}{c}{$y''$} &
\multicolumn{1}{c}{$\Delta x$} &
\multicolumn{1}{c}{$\Delta y$} &
\multicolumn{1}{c}{$I$} &
\multicolumn{1}{c}{$(V$--$I$)} &
\multicolumn{1}{c}{\scriptsize Type/ref} \\
\noalign{\smallskip} \hline \noalign{\smallskip}
     1 &   257.9   & --11.4   &   0.6   &  --0.9  &  15.56   &  0.54
&{\scriptsize ab  2} \\
     2 &   --40.5   &  29.3   &  --0.8   &   0.9  &  15.22   &  0.97
&{\scriptsize n,bl.  1,2} \\
     3 &   --42.6   &  94.4   &   0.6   &   0.6  &  15.61   &  0.32
&{\scriptsize c  2} \\
     4 &    25.7   &  35.6   &  --1.0   &   0.2  &  15.88   &  0.67
&{\scriptsize ab  2} \\
     5 &    41.3   &  40.2   &   0.0   &   1.0  &  15.35   &  0.49
&{\scriptsize ab  2} \\
     6 &   --74.0   &  --8.3   &  --0.3   &   0.1  &  15.81   &  0.68
&{\scriptsize ab  2} \\
     7 &     3.4   &--110.2   &  --0.4   &   0.2  &  15.45   &  0.49
&{\scriptsize ab  2} \\
     8 &    29.6   &  25.5   &  --0.7   &  --0.8  &  15.60   &  0.49
&{\scriptsize ab  2} \\
     9 &   --56.7   &  49.5   &  --1.1   &   0.0  &  13.59   &  0.24
&{\scriptsize lP,n  1,2} \\
    10 &    45.6   &--197.4   &   1.2   &   0.8  &  15.90   &  0.60
&{\scriptsize ab  2} \\
    11 &    67.0   &--136.8   &  --1.0   &  --0.2  &  15.72   &  0.66
&{\scriptsize ab  2} \\
    12 &   --77.2   & --45.9   &   0.2   &  --0.1  &  15.32   &  0.35
&{\scriptsize ab  2} \\
    13 &     5.4   &  45.9   &   0.6   &   0.1  &  15.70   &  0.20
&{\scriptsize c 2} \\
    14 &    74.6   &  16.1   &   0.4   &  --0.1  &  14.85   &  0.60
&{\scriptsize ab  2} \\
    15 &    32.5   &  51.2   &  --0.5   &  --0.2  &  15.49   &  0.51
&{\scriptsize ab  2} \\
    16 &    67.3   &  --0.7   &  --0.3   &  --0.3  &  15.98   &  0.45
&{\scriptsize ab  2} \\
    17 &   --42.2   & --55.5   &   0.2   &   0.5  &  15.55   &  0.66
&{\scriptsize ab  2} \\
    18 &    45.5   & 158.7   &   0.5   &   0.3  &  15.75   &  0.32
&{\scriptsize c 2} \\
    19 &    26.0   & --32.6   &  --1.0   &  --0.4  &  15.67   &  0.43
&{\scriptsize c  2} \\
    20 &   --10.5   & --23.0   &   0.5   &   0.0  &  15.69   &  0.57
&{\scriptsize ab  2} \\
    21 &   --58.7   &  61.8   &  --0.3   &   0.2  &  15.90   &  0.28
&{\scriptsize c  2} \\
    22 &   131.0   & 108.7   &   0.0   &  --0.7  &  15.47   &  0.55
&{\scriptsize ab, bl.  3} \\
    23 &   110.0   & --57.6   &  --1.0   &  --0.4  &  15.78   &  0.23
&{\scriptsize c  3} \\
    24 &   144.5   & --96.5   &  \multicolumn{4}{c}{saturated}
&{\scriptsize 183?  3} \\
    25 &  --131.1   &--274.6   &   2.1   &   0.6  &  14.86   &  0.79
&{\scriptsize c  3} \\
    26 &  --123.9   & 106.7   &  --0.1   &   0.3  &  15.68   &  0.41
&{\scriptsize fld.   3} \\
    99 &   --84.4   & 106.2   &   0.4   &  --0.2  &  15.51   &  0.63
&{\scriptsize n  3} \\
\noalign{\smallskip} \hline
\end{tabular}
\end{flushleft}
\end{scriptsize}
\end{table}

Extensive studies of NGC~1851 variables are found in Liller (1975),
Wehlau et al. (1978, WLDC) and Wehlau et al. (1982); on the basis of the
value $N_{c}$/$N_{ab}$ = 0.47, the cluster has been classified
as Oosterhoff type I.

Using the RR~Lyr coordinates given by these authors and by Fourcade \&
Laborde (1966), our catalogue has been searched in order to identify
the variables.  Twenty-six out of 27 variable stars have been successfully
found, and these are reported in Table~\ref{rr_lyrae}: in this table, $N$
is the identification number used in the literature, $x$ and $y$ are
the coordinates of our stars (in arcsec) and ``type'' is the RR~Lyr
type given by Wehlau et al. (1978, 1982) and Liller (1975). The frame
of reference is centered at ($\alpha$,$\delta$; eq.) = (5$^{\rm h}$
12$^{\rm m}$.4, --40$^\circ$ 05$'$; B1950.0).  The transformation from our
frame of reference to WLDC is:

\[
\left\{
\begin{array}{ll}
x_{\rm W} = -528.6 + x + 0.0179 \cdot y \\
y_{\rm W} = - 499.1 - 0.0179 \cdot x + y
\end{array}
\right.
\]

\noindent There is a rotation angle of $\simeq$ 1.03 deg; also, the
scale factor 0.44 must be applied to $x_{\rm W}$ and $y_{\rm W}$; the
residuals $\Delta x$ and $\Delta y$ in the coordinates after the
transformation are also reported in Table~\ref{rr_lyrae}.  Column 6
and 7 of Table~\ref{rr_lyrae} give our determination of $I$ and ($V$--$I$);
the mean magnitudes, colours, and periods of these RR~Lyr are listed in
WLDC78 and Wehlau et al. (1982).  In this table, actually 22 are true
RR Lyr variables (15 ab and 7 c type); stars 2, 9 and 99 are thought
to be non--variables (and actually do not scatter in
Fig. 17), star 26 could be a W UMa field star, and
finally star 24 could be a long--period variable. This latter star has
not been measured because it is located at the RGB tip (cf. W92), and
 is saturated in our images. Star 9 is a very red star below the RGB
tip: it is almost saturated in $I$, so it has an artificially blue colour
and for this reason it has not been plotted in our CMD.

Our photometry of the RR Lyr in Table~\ref{rr_lyrae} gives a mean
magnitude $<$V$>_{\rm RR}$ = 16.08 $\pm$ 0.06 mag, where the error
reflects the intrinsic dispersion of their distribution, which
corresponds to $<M_{V}({\rm RR})> = 0.58
\pm 0.20$ mag with the assumed distance modulus and absorption
(Sect.~\ref{abscalibration}).
Therefore, RR~Lyr
stars are 0.12 mag brighter than the ZAHB at $V_{\rm HB} = 16.2 \pm 0.03$
mag (cf. Sect.~\ref{abscalibration}).  Using the relation of Carney et
al. (1992) a value $<$V$>_{\rm RR}$ = 16.11 mag is predicted, well inside the
1-sigma uncertainty. 
The internal uncertainty in the RR~Lyr colours can be estimated from
Table 11, which gives $\leq$0.04~mag for $V$$<$16.74 mag. Adding
the calibration uncertainty, we obtain a total error of 0.06~mag.
Finally, we can also give for the first time the mean
$I$ magnitude of the RR Lyraes: we find $<$$I$$>_{\rm RR}$ = 15.59 $\pm$ 0.06 mag,
corresponding to $<M_{I}({\rm RR})> = 0.12
\pm 0.20$ mag, assuming $A_{I} = 0.03$ and $(m-M) = 15.44 \pm 0.2$.

\subsection{New RR Lyrae candidates} \label{sec_nuoveRR}

\begin{figure}[t]

\psfig{figure=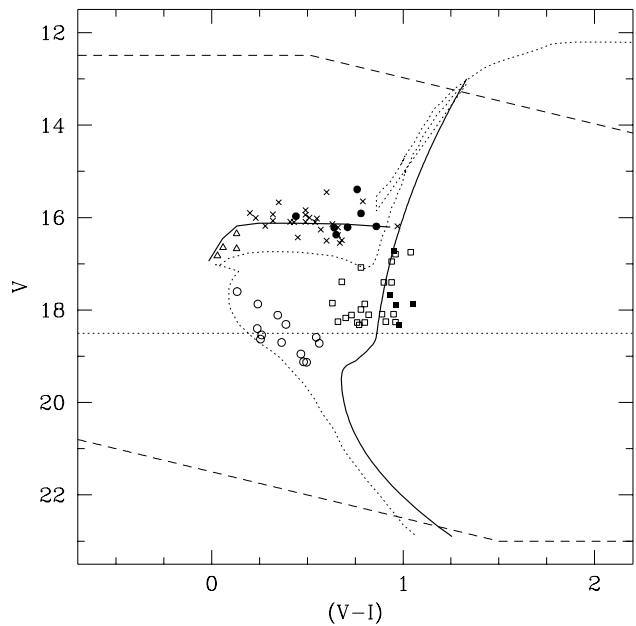,width=8.8cm}

\caption[]{
The location of known and suspected variables in the NGC~1851 CMD: the
different symbols refer to RR~Lyr stars (crosses), the candidate RR Lyraes
(full dots), and (c) other blue objects (open triangles) and red objects
(open squares and filled squares) which are discussed in
Appendix A.2. Also plotted are the blue straggler stars (open dots):
these stars are fainter than the limit imposed by a 2~Gyr, $Z$=0.001 isochrone
(dotted line), which has a TO mass $m_{\rm TO} \simeq 1.5\,M_\odot$. The
solid line represents the fiducial sequence of NGC~1851 and the horizontal
dotted line represents the fainter limit of the W92 photometry. See also
discussion in Appendix A.2. }
\label{var_cmd}
\end{figure}

\begin{table}[t]
\begin{flushleft}
\caption[]{Positions and photometry of variable candidates.}
\label{new_variables}
\begin{tabular}{rrrrrrrr}
\hline \hline \noalign{\smallskip}
\multicolumn{1}{c}{$n$} &
\multicolumn{1}{c}{$x''$} &
\multicolumn{1}{c}{$y''$} &
\multicolumn{1}{c}{$V$} &
\multicolumn{1}{c}{($V$--$I$)} &
\multicolumn{1}{c}{$\Delta V_{\rm W}$} &
\multicolumn{1}{c}{$\Delta V_{\rm RR}$} &
\multicolumn{1}{c}{Type} \\
\noalign{\smallskip} \hline\noalign{\smallskip}
     1  & --38.5  & --28.8  &   15.39  &   0.76  &  --0.23  &  --0.66 & ab \\
     2  &  --9.2  &  32.9  &   16.22  &   0.64  &   0.38  &   0.17 & ab \\
     3  & --31.6  & --11.0  &   16.37  &   0.65  &   0.35  &   0.32 & ab \\
     4  & --18.9  &  38.1  &   16.21  &   0.71  &   0.15  &   0.16 & ab \\
     5  & --15.6  & --17.8  &   15.91  &   0.78  &  --0.18  &  --0.14 & ab \\
     6  &  --5.2  & --53.1  &   15.97  &   0.44  &  --0.35  &  --0.08 & c \\
     7  & --12.1  & --20.2  &   16.19  &   0.86  &  --0.14  &   0.14 & ab \\
\noalign{\smallskip} \hline
\end{tabular}
\end{flushleft}
\end{table}

When our photometry is compared with that of W92, it can be seen
(Fig. 17) that a number of stars at the level of the HB are
located outside the ``3-$\sigma$'' lines. These stars have been extracted
from the photometric catalogue and inspected both on the $V$
and $I$ images, in order to remove non-stellar objects. A ``quality'' parameter
has been devised, including remarks about the FWHM, the presence of
neighbours, the shape and the blending: the stars that passed these criteria
are listed in Table~\ref{new_variables}. From left to right, the columns
report an identification number, the coordinates in the WLDC
frame of reference, the magnitude and colour in our photometry, the magnitude
difference with respect to the W92 values and the mean RR~Lyr value, and
finally the estimated variable type.
These stars
are also plotted in Fig.~\ref{var_cmd},
where different symbols are used to identify (a) the known RR~Lyr
stars ({\it crosses}) discussed in the previous section and (b) the
new RR~Lyr candidates ({\it full dots}).

From the location in Fig.~\ref{var_cmd}, six of the seven RR Lyr candidates
can be assigned to the $ab$ type and one to the $c$ type: this would bring
the ratio $N_c$/$N_{ab}$ to the value $0.38 \pm 0.22$, in better agreement
with the value found for the other clusters of comparable metallicity.  
In fact, the catalogue of Castellani \& Quarta (1987) can be used to
calculate the expected $N_{\rm c}$/$N_{\rm ab}$ ratio. There are six
clusters with [Fe/H]$\sim -1.3$ in the compilation. If we exclude
NGC~1851 itself, NGC~362 (which has no $c$~type RR~Lyr) and NGC~6864
(which has only six variables,  whose ratio also happens to be too uncertain), a
weighted mean of the three remaining clusters yields a value $N_{\rm
c}$/$N_{\rm ab} = 0.23 \pm 0.09$.  For comparison, Smith (1995) gives
a value $N_{\rm c}$/$N_{\rm ab} = 0.20$ as representative of the Oo~I
clusters. The ratio of NGC~1851 is therefore 2.6 sigma higher than the
value found for the clusters of similar metallicity.

A determination of the light curves for the
candidate variables is required in order to confirm this result in the 
N$_c$/N$_{ab}$ ratio of the RR Lyr variables in NGC 1851.

\subsection{The blue straggler stars}\label{bss}

\begin{table}[t]
\caption[]{Coordinates and photometry of the candidate blue straggler
stars. The $x$ and $y$ data are expressed in arcsec in the Wehlau et
al. (1978) frame of reference.
}
\label{cooBSS} 
\begin{tabular}{rrrrr}
\noalign{\smallskip}
\hline\hline
\noalign{\smallskip}
\multicolumn{1}{c}{$N$} &
\multicolumn{1}{c}{$x''$} &
\multicolumn{1}{c}{$y''$} &
\multicolumn{1}{c}{($V$--$I$)} &
\multicolumn{1}{c}{$V$} \\
\noalign{\smallskip}
\hline
\noalign{\smallskip}
 1  &  107.4 & --194.0 & 0.13 & 17.60  \\ 
 2  & --124.9 &    2.6 & 0.38 & 18.31  \\ 
 3  & --156.6 &  114.6 & 0.46 & 18.95  \\ 
 4  & --110.6 &  127.7 & 0.23 & 18.40  \\ 
 5  &  186.7 &  189.8 & 0.54 & 18.59  \\ 
 6  &  131.9 &  194.3 & 0.25 & 18.63  \\ 
 7  &  --83.1 & --152.8 & 0.34 & 18.11  \\ 
 8  &  160.9 &  --46.6 & 0.36 & 18.70  \\ 
 9  &  222.4 &   44.8 & 0.56 & 18.72  \\ 
 10 &    6.5 & --164.0 & 0.49 & 19.13  \\ 
 11 &  243.1 &   93.6 & 0.26 & 18.54  \\ 
 12 & --132.1 &  427.8 & 0.24 & 17.87  \\ 
 13 & --292.2 & --148.9 & 0.48 & 19.12  \\ 
\noalign{\smallskip}
\hline
\end{tabular}
\end{table}

Thirteen blue straggler stars have been identified in the photometric
catalogue in the region $r>80$ arcsec (the crowding did not allow a 
clear identification of BS stars closer to the cluster center) and they are
listed in Table~\ref{cooBSS}. The columns are, from left to right, the
identification number, the positions in the WLDC frame of reference, the
colour and magnitude in our photometric system.

Blue straggler stars are objects which are located in the
region between the TO and the blue tail of the HB and are
distributed along a sequence normally filled by MS stars younger than
those of a GC. Bolte et al. (1993) and Ferraro et al. (1995), among
others, have studied the BS stars frequency in GGCs, and Ferraro et
al. (1995) found that clusters which are more massive and denser have
more BS stars with respect to clusters of lower mass and density.

BSs are thought to be the result of the merging (by coalescence of binaries
or collision) of two main-sequence stars. If we define $m_{\rm BSS}$ the
mass of any BS stars and $m_{\rm TO}$ the mass of a TO star, it should
always be $m_{\rm BSS} < 2\, m_{\rm TO}$.
The BS star sample in NGC~1851 is compatible with this statement, as
can be seen in Fig.~\ref{var_cmd}: all the 13 identified BS stars are less
luminous than the limit set by a $Z = 0.001$, 2~Gyr isochrone, which
has a TO mass $\simeq 1.5 \, M_\odot$; by comparison, the TO stars of
a 15~Gyr isochrone have a mass $\simeq 0.8 \, M_\odot$.

The BS radial distribution has been compared with the distribution of other
reference stars, i.e.  the SGB stars with the same magnitude as the BS.  We
found that the BS of NGC 1851 are {\sl less} concentrated than the SGB
stars, with a confidence level of 96\%. This result might seem to
contradict the usual statement (Bailyn 1995) that the BS are more
concentrated than the other cluster stars. Indeed, this is not generally
true, as, at least in one GC (M3), the BS show a bimodal distribution (more
concentrated than the other cluster stars in the core and less concentrated
in the outer envelope, Ferraro et al. 1997). The NGC~1851 BS population could
have a similar distribution (Piotto et al. private comm.).

Concerning the stars evolved from the BS group, we expect to recover
these in a clump above the RHB region around $V\simeq 15.3$ mag and
($V$--$I$)$\simeq 0.8$ (cf. the 2~Gyr isochrone in Fig.~\ref{var_cmd},
dotted line). Indeed, this region is populated by a significant group
of stars, but (a) these stars are compatible with post-HB evolution
(cf. Sect.~\ref{sect_shb}) and (b) given the small number of BS stars
their descendants are too few, rendering their recognition impossible.

Using the same method of Sect.~\ref{sec_nuoveRR}, no BS stars are found
to be variable: we flag as variable candidates those stars which
scatter more than $3\sigma$ with respect to the W92 values, and for
all BS stars we find $| V - V_{\rm W92} | < 0.2$ mag, which is the
resulting threshold for 17 mag $< V < $19 mag. More than 50\% of contact
binaries and 70\% of dwarf Cepheids have light amplitudes greater than
0.2~mag (Mateo 1993), so if we consider the already cited relative
frequencies of the two types of variables, we expect that, of our 10 BS
stars, $\sim 1$ and $\sim 0.2$ should be pulsating stars or contact
binaries, respectively. The null result for NGC~1851 is therefore
compatible with the observed frequency of variable BS stars in other
clusters, if we consider the great uncertainties that affect these
numbers.

\section{The horizontal branch} \label{thehb}

Starting with S81, all previous photometric studies emphasize the
anomalous morphology of NGC~1851's horizontal branch, showing both a
red clump and a blue tail (Fig.~\ref{complete_cmd}).  Since Sandage \&
Wildey (1961), it is well-known that a  relation exists
between the metallicity of a GC and its HB morphology, with clusters
with higher [Fe/H] having redder HBs. Red clumps are typical of 
metal-rich GGCs, such as 47 Tuc, and blue tails are typical of metal--poor
clusters, such as M30.  The bimodality of the NGC~1851 HB is an exception
to this scheme. Neither it is an example of a {\sl second parameter}
cluster: in such a case, it would show a HB morphology typical of a
GGC with a {\sl different} metallicity, but would otherwise be
normal. NGC~1851's HB is peculiar, then, in the sense that it looks like
a mixture of two HBs of different metallicities.

The deepness and richness of the present sample allows a new detailed
comparison with up-to-date stellar models, and for the first time, the
full radial coverage gives the opportunity of investigating the HB
morphology as a function of the distance from the cluster center.

\begin{figure}[t]

\psfig{figure=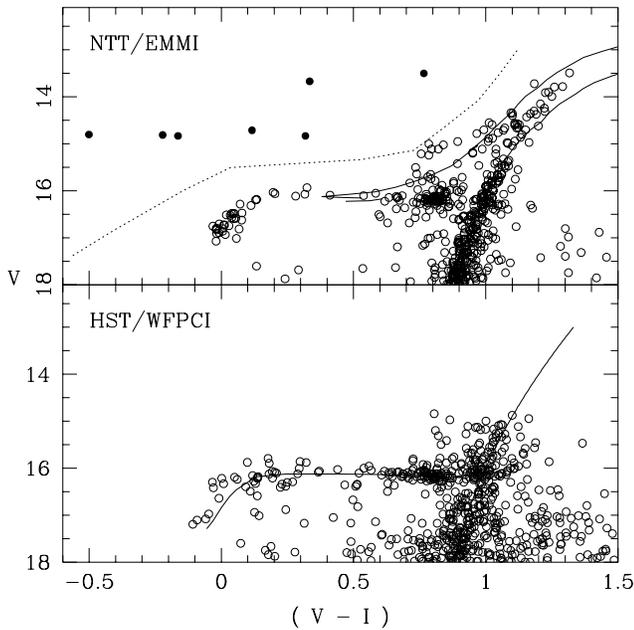,width=8.8cm}

\caption[]{The horizontal branch of NGC~1851. {\it Upper panel}:
colour--magnitude diagram from the NTT/EMMI data (22 $\times$ 22 arcmin$^2$
total area). In order to better display the main branches, only stars with an
{\sc allstar} parameter $\chi$ $\leq$ 1.5 have been plotted.
Saturated stars and suspected blends have been removed from the CMD area
bluer than the dotted line.  A 15.1~Gyr,
$Z$ = 0.001 isochrone from Bertelli et al.  (1994) is also plotted:  it shows
that, using a set of parameters valid for NGC~1851, standard evolutionary
models are not able to predict the observed blue HB tail.  {\it Lower panel}:
colour--magnitude diagram from the {\it HST}/WFPCI data (central $\sim$~30$''$).  In
this case, saturation prevents the study of possible supra-HB stars.
In both diagrams, the bimodal appearance of the cluster HB (the red clump
plus the blue tail) can be clearly seen.  }
\label{cm_supra_hst}
\end{figure}

Figure~\ref{cm_supra_hst} shows an enlargement of the HB region of the CMDs of
NGC~1851. The peculiar morphology of the HB is apparent: despite the fact
that NGC 1851 has a [Fe/H]=--1.28, there is a very well defined red clump
(0.7 $<$ ($V$--$I$) $<$ 0.9), followed by the RR~Lyr gap at 0.2 $<$ ($V$--$I$) $<$
0.8, and a blue horizontal branch with a long blue tail ($\simeq$ 1 mag)
which extends down to $V$ $\simeq$ 17.3 mag.

There are no gaps in the distribution of the BHB stars, as already remarked
by W92. The observed fractions of the HB populations are (BHB:RR:RHB) =
(0.25:0.08:0.67), which agree with W92 values (0.27:0.09:0.64).

Lee et al. (1988) claimed that, in the case of
NGC~1851, HB bimodality could be reproduced by taking into account evolved
off-ZAHB stars, but two facts
contradict this explanation:
first, Lee (1992) predicts population ratios with values (BHB:RR:RHB) =
(0.81:0.14:0.05), which do not match the observed values (cf. above); 
secondly, as noted by W92, the lifetimes for the tracks of Lee \&
Demarque (1990)
imply that He-burning stars spend most of their life near the ZAHB.

\subsection{Synthetic horizontal branches} \label{synthHB}

\begin{figure}[t]
\psfig{figure=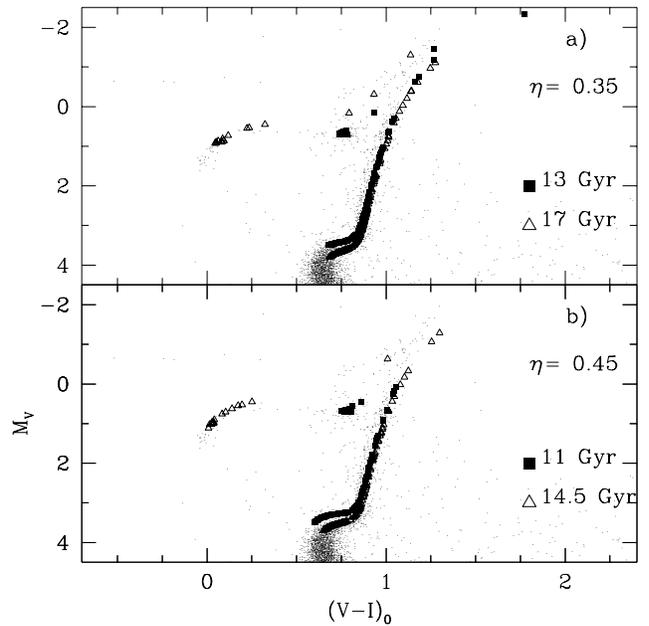,width=8.8cm}
\caption[]{Synthetic CMDs obtained for $\eta$ = 0.35
and ages of 13 and 17 Gyr (Panel $a$) and for $\eta$ = 0.45 and ages
of 11 and 14.5 Gyr (Panel $b$) superimposed on the observed CMD for
NGC~1851.  The models have been calculated adopting the isochrones by
Bertelli et al. (1994) with metallicity $Z$=0.001 and assuming an
instantaneous initial burst of star formation; no dispersion in mass-loss
 efficiency has been introduced.  The colour-excess is assumed to
be $E(V-I)$ = 0.02 and the apparent distance modulus $(m-M)$ = 15.45.
({\it a}) case $\eta$=0.35: the 13 Gyr old simulation can fit the red
HB stars but  does not reproduce the blue part of the HB. Conversely
the 17 Gyr is able to fit only the blue HB stars.  ({\it b}) case
$\eta$=0.45: in this case the two ages for which the red and blue
parts of the HB can be independently fit are 11 and 14.5 Gyr,
respectively.}
\label{synthetic_1}
\end{figure}

\begin{figure}[t]
\psfig{figure=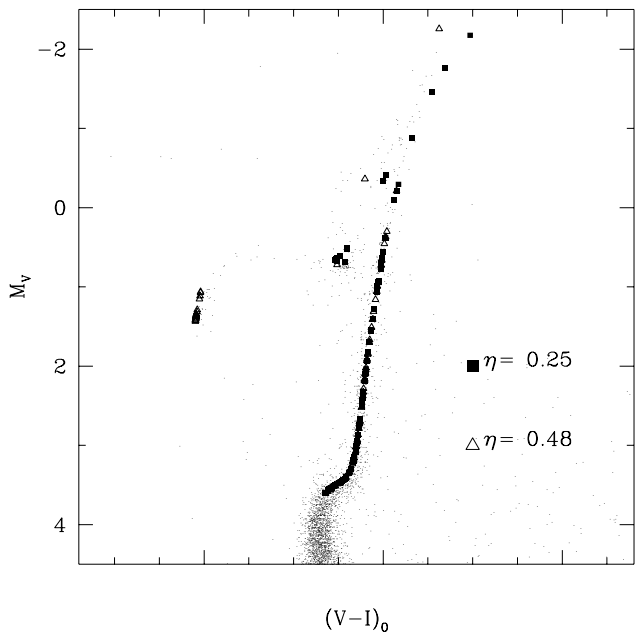,width=8.8cm}
\caption[]{As in Fig.
\ref{synthetic_1}
but for the case of variable
$\eta$.
The age at which the difference in luminosity from the turn-off and the HB
is reproduced corresponds to an age of 15 Gyr. Assuming this age, it
turns out that the red HB is reproduced by a mass-loss efficiency
of $\eta$ = 0.25 while the blue HB by $\eta$ = 0.48.}
\label{synthetic_2}
\end{figure}

Standard evolutionary models are not able to reproduce the observed
distribution of stars along the HB: the solid line in the upper panel
of Fig.~\ref{cm_supra_hst} represents a 15.1 Gyr, $Z = 0.001$ isochrone
from Bertelli et al.  (1994) which better fits the luminosity
difference between the HB and the TO levels.  The theoretical track
(which adopts a standard value for the RGB mass-loss efficiency $\eta
= 0.35$) predicts only an intermediate--colour HB, as is expected
from the cluster's metallicity.  The parameter $\eta$ is the one used
in the Reimers (1975) formula ($\dot{M} = 4 \cdot 10^{-13}\,
\eta\,L/(g\,R)$~$M_{\odot}$~yr$^{-1}$), which is widely employed to
model stellar mass loss along the RGB phase.  The HB initial mass is
0.65 $M_\odot$. A certain dispersion in colour along the HB is expected
due to the possible scatter in the mass-loss rate. The resulting
dispersions in the HB mass is estimated at $\sim 0.01 \, M_\odot$ (Iben
\& Rood 1970), which is not enough to produce masses as low as
0.60 $M_\odot$ or as high as 0.70 $M_\odot$, as needed to match the
observed HB colours (cf. Sect.~\ref{sect_shb}).

A simple way out of this discrepancy is to suppose that {\sl two}
populations of stars actually be present on the HB, and that some
differences in their basic parameters cause their colour ranges to be
different. Among these basic parameters, the possibility that age or
mass loss could be responsible for the HB bimodality has first been
explored.  The observed CMD has been compared with the synthetic
diagrams, specifically generated by Monte Carlo simulations.  The
simulated CMDs are based on the assumptions that:

\begin{enumerate}
\item the evolutionary isochrones are from Bertelli et al. (1994)
and are based on the models with chemical composition
[$Z, Y$] = [0.001, 0.230]
(the conversion from theoretical to observational plane
is made by convolving the library of stellar spectra by Kurucz, 1992; the
theoretical colours have been specifically calculated for this paper);

\item the initial mass function is a Salpeter law with $ x =1.35 $; and

\item the star formation is assumed to take place in an almost instantaneous
initial burst.

\end{enumerate}

\noindent
We did not include the observational errors and no attempt has been made to
reproduce the number of stars in a given interval of magnitude or
evolutionary phase.  The comparison of the synthetic diagrams with the
observed CMD is shown in Fig.~\ref{synthetic_1}. It is evident that {\sl a
unique age is not able to fit the red and blue part of the HB at the same
time}.  Indeed, the HB of NGC~1851 would require that two populations with an
age difference of $\sim$ 4 Gyr are present in the cluster, which is quite
difficult to accept.
Furthermore, Fig.~\ref{synthetic_1} shows that in this case we would expect
a spread in the SGB luminosity much larger than the photometric errors,
which is not detected.

However, if we assume that the stars in NGC~1851 are 15 Gyr old
(as is suggested by the difference between the HB and TO luminosities, as
previously discussed), we need a bimodal mass loss to reproduce the HB
morphology, as shown in Fig.~\ref{synthetic_2}: stars with higher and lower
than standard $\eta$ values are needed in order to populate both the BHB
tail and the RHB. Of course, a supposed dispersion in the two adopted $\eta$
values could reproduce the observed distribution of stars along
the entire HB sequence.

A bimodal distribution of the other possible second parameters is excluded by
what it is currently known about helium in GCs (Buzzoni et al. 1983) or the
distribution of CNO elements in NGC~1851
(Armandroff \& Zinn 1988).
Only two stellar groups with two different mass loss efficiencies during
the RGB phase can correctly reproduce both the red and the blue tail of the HB
of NGC~1851.

\subsection{Environment and stellar evolution}

\begin{figure}[t]

\psfig{figure=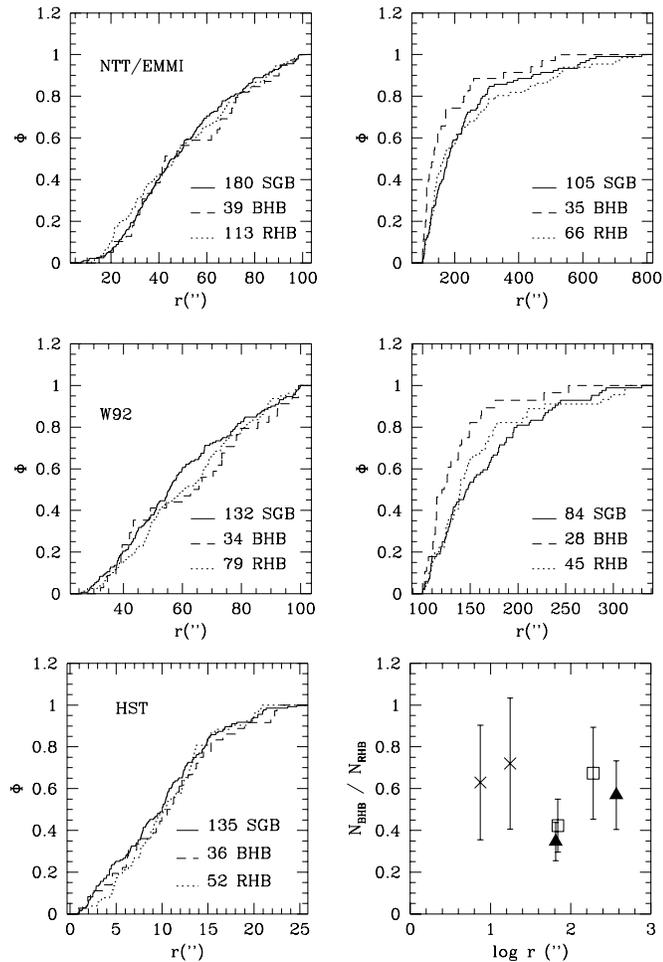,width=8.8cm,bbllx=160pt,bblly=340pt,bburx=410pt,bbury=710pt}

\caption[]{
The upper and central panels show the cumulative distributions for
three samples of stars for the EMMI data and for the W92 data,
respectively.  The same distributions are shown for the {\it HST} sample in
the lower left panel, while the ratio $N_{\rm BHB}/N_{\rm RHB}$ vs. radius is
plotted in the lower left panel. Crosses indicate the {\it HST} sample, open
squares indicate the W92 sample and filled triangles indicate the EMMI
sample.
}
\label{ks_tests}
\end{figure}

In order to define the boundaries of the
HB regions in an unbiased manner, the HB has  been linearized:
a fit has been drawn by eye through the points;
then each $V $ magnitude has been
replaced by the difference $V$--$V_{\rm fit}$, where $V_{\rm fit}$ is the
magnitude of the line at the same colour.
Finally, the
distributions in colour and magnitude have been inspected,
and the limits for the different groups of stars have been selected
at the color where the histograms showed the maximum gradient;
the same procedure has been
followed for the W92 data, which have been included in the present analysis.

We have checked for possible gradients in the distribution of the
SGB--RGB--HB stars. First of all, we note that the giants in the
brightest 2 magnitudes of the RGB are saturated, both in the NTT/EMMI
and in the {\it HST} sample. This prevents a meaningful check of
possible demise of red giants in the inner core, as found in many
other high-density clusters, and a study of the radial trend of the
mean cluster colour to check whether the UV gradient found by Dupree et
al. (1979), for the same cluster, extends to the optical bands.

In the SGB star sample we have included all the stars with 15.5 mag $\leq V
\leq 17$  mag and $0.9 \leq (V-I) \leq 1.1$ for the NTT/EMMI and {\it HST} samples, or
$0.75 \leq (B-V) \leq 1.0$ for the W92 sample.  Figure~\ref{ks_tests}
shows the cumulative distributions in radius of BHB, RHB and SGB
stars, for the three data sets considered.  There is some evidence
that the behaviour of BHB stars with respect to SGB stars is different
from that of RHB stars: Up to 80\arcsec~~ the distributions show
similar trends, whereas outside this limit BHB stars seem more
concentrated than SGB/RHB stars.

 %
 % Macro: /home/ivo/tex/n1851/hb/new_hb.mac --> many_profs
 %
 %

\begin{figure}[t]
\psfig{figure=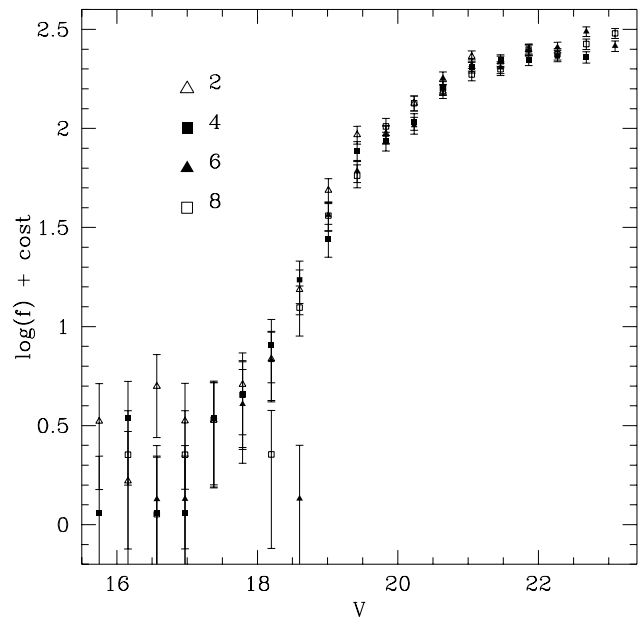,width=8.8cm}
\caption[]{The corrected luminosity functions from fields 2, 4, 6 and 8;
the LFs have been normalized to $10^3$ objects in the range 19 mag $\leq V \leq
22$  mag.\label{4lf}}
\end{figure}

We run also a two-population Kolmogorov--Smirnov test.  The test gives
no statistically significant results when it is applied over the whole
radial range; the significance is higher when the distributions are
analysed in two radial ranges.  The probabilites are computed in the
$r <$ 100\arcsec~ and $r >$ 100\arcsec~ ranges.  Up to 100\arcsec~ the
probability that RHB, BHB and SGB stars are taken from the same
distributions are quite high for all the data sets (from $\sim$~50~\%
to $\sim$~80~\%). For $r > 100$~\arcsec~, the same test gives low
probabilities ($\sim2\%$) for similar distributions when comparing the
BHB with the SGB stars, and high probabilities for similar
distributions when comparing the RHB with the SGB stars.  The BHB
stars appear to be more concentrated than the RHB and the SGB stars.

From this test, we cannot conclude that there is a significant trend,
even in the outer part of the cluster, although this possibility cannot
be excluded: see Piotto et al. (1988) and Djorgovski \& Piotto (1993)
for a discussion of the limits in applying these statistical tests for
checking population gradients.  In any case, a gradient in the HB
stars present only at large distance ($r \gg r_c$) cannot be easily
interpreted from the dynamical point of view. However, this might not
be the only anomaly in the distribution of the stars of NGC 1851. Also the
BSs in the outer envelope have a distribution different from that
in the inner core (cf. Sect.~\ref{bss} and Piotto et al. 1997).
It is tempting to imagine that the two anomalies might originate from
the (still unknown) dynamical mechanism.

We still remain with the UV gradient found in the cluster core
by Dupree et al. (1979). It will be of particular
interest to check the distribution of the RGB stars, even if it might
be possible that the supra-HB stars in Fig.~\ref{complete_cmd},
which are found in the most inner region are, at least in part,
responsible for the UV gradient.

The presently available data do not show any direct evidence that the
higher mass loss needed to explain the bimodal-HB of NGC 1851 is
related to the very high density inner environment of this cluster,
though the anomalous distribution of the BHB and BS stars still need
an interpretation. Finally, we note that the presence of an anomalous
blue-HB well fit the scenario depicted by Fusi Pecci et al. (1993),
showing a correlation between the presence of a high-density core and
a blue-HB tail.

\section{Luminosity and mass functions} \label{lfmf}

In order to define a sample of stars belonging to the MS--RGB branch, we
selected those objects falling within 2~$\sigma$ from the linearized
ridge line.
The sequence widths in Table \ref{fid_points}, have been fitted with the
law
\[
\sigma = \exp\,(a\,V + b),
\]
where the parameters turn out to be $a = 0.36$ and $b = -9.85$, and the
colour of the ridge line has been fitted with a spline and subtracted star
by star according to $V$ magnitude in order to linearize the sequence. These
selected samples will be referred to as ``$2\sigma$'' data sets.

The external field (cf. Fig.~\ref{background_cmd}) has been used to estimate
the background/fore\-ground contamination: 19 stars are counted for $V\leq$
22 mag (within the estimated 100\% completeness limit) over an area of 71
arcmin$^2$ (``$2\sigma$'' sample), while 13224 stars are recovered in the
same $V$ range for the total CMD (``$2\sigma$'' sample). Given an area of
484 arcmin$^2$ for the CMD, this makes a $\sim$~1\% contamination level,
which can be safely neglected.

The crowding experiments have been run only for the external fields 2, 4, 6
and 8, and for the central field 5: the completeness correction is then
possible only for these fields. Due to the lower surface density of stars,
the deepest photometry is that of the external fields, and these will be
discussed first. The single corrected LFs of the four external fields
($2\sigma$ samples) are shown in Fig.~\ref{4lf}, down to the 50\%
completeness limit; the LFs have been normalized to $10^3$ stars in the
range 19 mag $\leq V \leq 22$ mag, so they can be directly compared. The
figure shows that the four LFs well agree within the error bars, and that
the photometry goes fainter and fainter in field order 2, 4, 6 and 8. As the
overall seeing and distance from the cluster center are the same for the
four fields, the different limiting magnitudes are due to the different
exposure times, which are comparable for fields 6 and 8, and $\sim$ 0.8 and
0.6 times shorter for fields 4 and 2, respectively. This is consistently
confirmed by the completeness curves: the 50\% level is reached at $V
\simeq 23$ mag for the fields 6 and 8, while it is $\sim$ 0.2 and 0.8 mag
brighter for fields 4 and 2, respectively. In order to have a reliable
estimate of the (external) mass function (MF) trend, the external LF has
been built as the sum of the single LFs from fields 6 and 8: this allows a
greater luminosity range, and hence mass range, to be probed.

\begin{tiny}
\begin{table}[t]
\caption[]{
Values of the mass--function slopes $x$ derived from the external and
the internal LF; the conversion to mass ranges has been calculated
with the listed model isochrones and ages with an assumed apparent
distance modulus $(m-M)$ = 15.5 and metallicity $Z$ = 0.001. Each
isochrone is labelled according to the numbers at the bottom of the
table.
}
\label{mass_index} 
\begin{tabular}{rrrrrr}
\noalign{\smallskip}
\hline\hline
\noalign{\smallskip}
\multicolumn{1}{c}{Ref.} &
\multicolumn{1}{c}{Age} &
\multicolumn{1}{c}{$x_{\rm ext}$} &
\multicolumn{1}{c}{$\Delta\,x$}&
\multicolumn{1}{c}{$x_{\rm int}$} &
\multicolumn{1}{c}{$\Delta\,x$}\\
\noalign{\smallskip}
\hline
\noalign{\smallskip}
1 &   15.0  &   0.78  & 0.17 & --0.12 & 0.26\\
2 &   15.1  &   0.38  & 0.16 & --0.66 & 0.14\\
3 &   15.0  &   0.17  & 0.16 &   0.25 & 0.47 \\
4 &   10.0  &   1.53  & 0.44 &  --   & -- \\
5 &   10.0  &   1.52  & 0.18 &   0.89 & 0.20 \\
\noalign{\smallskip}
\hline
\noalign{\smallskip}
1 & \multicolumn{5}{l}{Bergbush \& Vandenberg (1992) } \\
2 & \multicolumn{5}{l}{Bertelli et al. (1994)        } \\ 
3 & \multicolumn{5}{l}{Girardi et al. (1997)         } \\
4 & \multicolumn{5}{l}{Alexander et al. (1997)       } \\
5 & \multicolumn{5}{l}{D'Antona \& Mazzitelli (1996) } \\
\noalign{\smallskip}
\hline
\end{tabular}
\end{table}
\end{tiny}

\begin{figure}

\psfig{figure=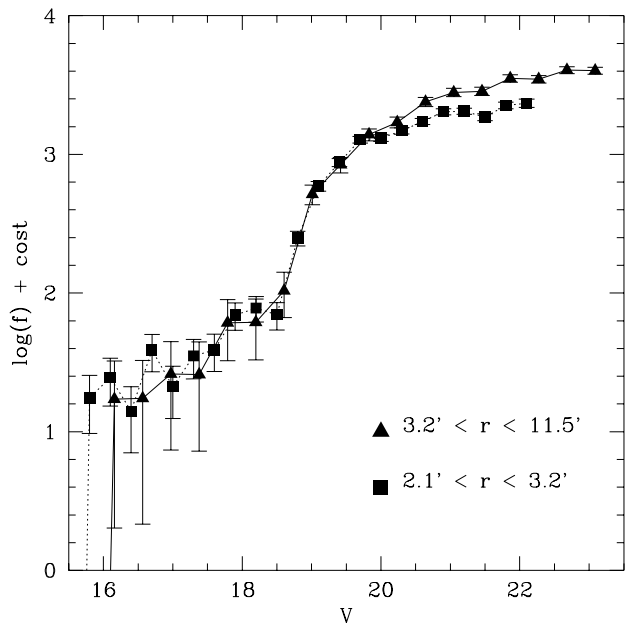,width=8.8cm}
\caption[]{Comparison of the NGC~1851 LF in two different radial ranges}
\label{interna_esterna}
\end{figure}

In the case of the central field, the output from the crowding tests reveals
that inside $\sim$~100 arcsec the completeness is so low ($< 50$\%, even
around the TO level) and completeness gradients are so steep that no
meaningful counts can be used. We then considered only stars outside this
limit and divided the sample into radial bins, such that each bin contained
the same number of objects, and computed the LF and the completeness curve
inside each bin.  
In order to go far enough inside to reveal
any radial difference in the LF, and deep enough with the photometry to
provide a large-magnitude baseline where the LFs can be compared, a
compromise must be reached. We chose to compute the internal LF in the
limits 120\arcsec~ $< r <$ 189\arcsec~, within which a magnitude $V \simeq$
22 mag can be reached at the 50\% completeness level.

The external LF has been computed in the radial interval 190\arcsec~ $< r <$
650\arcsec~, down to $V$ = 23.5 mag. The internal LF is shown in
Fig.~\ref{interna_esterna}, compared with the external one; the two LFs have
been normalized to 10$^3$ objects in the range 19 mag $\leq V \leq 20$ mag.
Although the magnitude interval is limited by completeness, in the $\sim 3$
common magnitudes below the TO, some sign of mass segregation is visible,
with the external LF steeper than the internal one.

\begin{figure}[t]
\psfig{figure=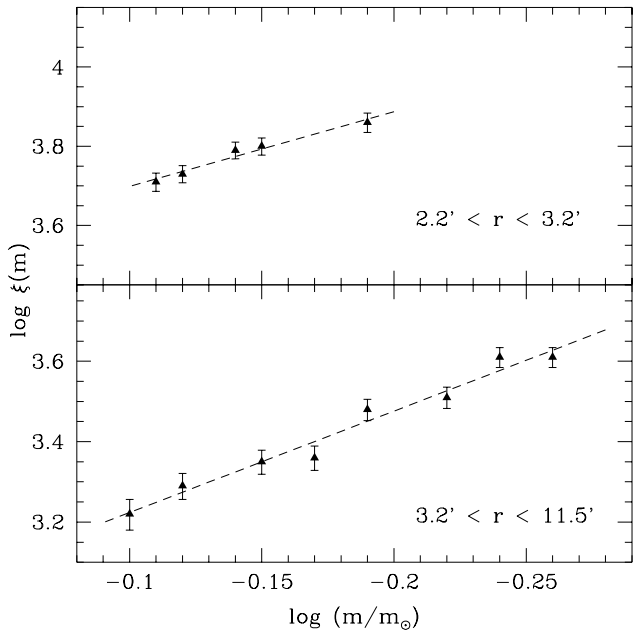,width=8.8cm}
\caption[]{Comparison of the NGC~1851 MF in two different radial ranges;
the transformation to mass has been computed by means of the theoretical
isochrones by D'Antona \& Mazzitelli (1996).}
\label{dos_dantona}
\end{figure}

It is now possible to convert the observed luminosities to masses in
order to obtain the mass--function slope, $x$, and to study the degree
of mass segregation.  The usual parameterization of the MF, $d\,N
\propto M^{-(1+x)} \, d\,M$, has been used.  Many theoretical
mass--luminosity relations (MLR) can be found in the literature: we
have no {\it a priori} reasons to favour one model with respect to
another, so the mass-function slope has been computed for all the
models listed in Table~\ref{mass_index}.  However, the
morphology of the observed MS is better reproduced by the Alexander et
al. (1997) and D'Antona \& Mazzitelli (1996, DM96) models; since the
Alexander et al. models do not reach bright enough magnitudes, we used
the DM96 models to transform luminosities into masses.

The choice of an MLR is very critical when deriving the mass--function. It
can be seen from Table~\ref{mass_index} that the slope $x$ varies within one
order of magnitude, going from the smaller values obtained from the Girardi
et al. (1997) models to the larger ones obtained from DM96 models. Past
investigations have already pointed out this problem (e.g. D'Antona \&
Mazzitelli 1983; Kroupa et al. 1993; Elson et al. 1995; Kroupa
\& Tout 1997, among others), which essentially bears on the fact that no
empirical MLRs are available for Population~II stars. A test of goodness for
the MLRs must therefore be restricted to the comparison to solar vicinity
data (e.g. Henry \& McCarthy 1993) and to an intercomparison between
different models or the CMDs for low metallicity relations. Such a test has
been made by Kroupa \& Tout (1997), and it turns out that, among the models
which cover GC-like metallicities, those by DM96 provide the best
mass--luminosity relations, which further justifies our choice.

The MFs are compared in Fig.~\ref{dos_dantona}. The difference between
the external and internal mass functions reflects the difference in
the LFs, and confirms that in the inner regions low-mass stars are
depleted with respect to high--mass stars. The figure also shows that,
in the mass range $0.5 < M/M_\odot < 0.8$ the MFs can be well fitted
by a power law.

The slope of the global mass function can now be obtained by correcting the
two observed mass functions for the effects of mass segregation.  As already
recalled, we lack the counts for the faint stars in the central regions, so
an observed global mass function cannot be extracted from our data. We have
calculated the mass segregation correction using multi-mass King--Michie
models, constrained by the density profile of the cluster.

\begin{figure}[t]
\psfig{figure=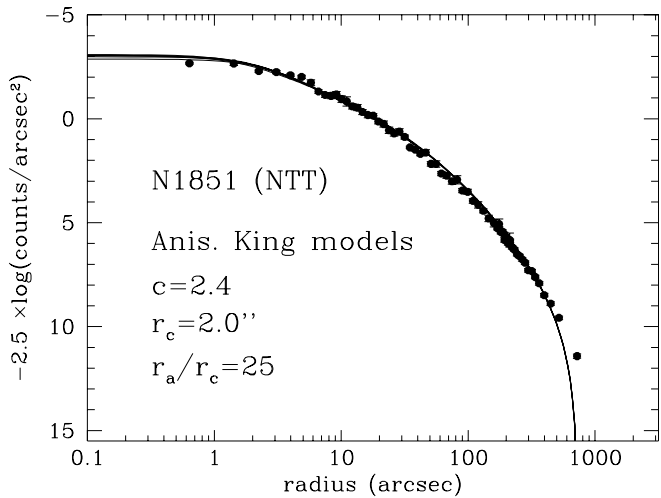,width=8.8cm,angle=-90,bbllx=90pt,bblly=80pt,bburx=540pt,bbury=680pt}
\caption[]{Closed dots: radial density profile of NGC~1851.
Continuous lines density profiles of the models fitting NGC~1851.
The parameters of the cluster models are also reported in the figure.}
\label{radprof}
\end{figure}

The radial profile of NGC~1851 was constructed in two ways:
(1) using the surface brightness profile for the internal part,
obtained from the shortest-exposure, non-saturated, best-seeing, $V$
image of the centre (image nr. 5, see Table~2); (2) with the star
counts of the ``$2 \, \sigma$'' sample for the external part of the
cluster (approximately outside $120''$).  The star counts were
completeness corrected down to $V$~=~22.5~mag and corrected for
background/foreground star contamination.  The overlapping region
between the two parts has been used to connect them.  The final radial
profile transformed in magnitude, with the central part rescaled to
the star counts, is shown in Fig.~\ref{radprof}.  The overall
behaviour is similar to the profile of NGC~1851 published by Trager
et al. (1995): it shows the same small depression at $\simeq8''$
present in the Trager et al. profile.  Our profile is a bit more
radially extended.

\begin{figure}[t]
\psfig{figure=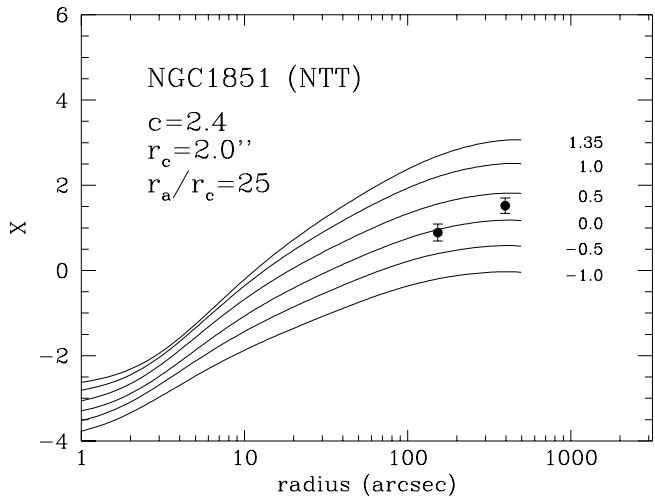,width=8.8cm,angle=-90,bbllx=90pt,bblly=80pt,bburx=540pt,bbury=680pt}
\caption[]{The filled circles represent the two slopes of the MFs of NGC~1851.
The continuous lines are the slope profile for different multi-mass
King models fitting the density profile.  The numbers on the right end
are the values of the global mass function used to build each model.}
\label{pryor}
\end{figure}

We have calculated the mass segregation correction as in Pryor et al.
(1986), following the recipe of Pryor et al. (1991) for the
construction of the multi--mass models.  We adopted a power-law MF, with
slope index $x$: this is justified by the fact that the two MFs of the
cluster follow fairly well a power law in the covered mass range,
$0.5<M/M_{\odot}<0.8$ (see Fig.\ref{dos_dantona}).  In the seven mass bins we
put stars starting from the turn-off mass down to 0.1 $M_{\odot}$,
plus a mass bin for the stellar remnants (in total eight mass bins).
Finally, in order to build the mass-segregation curves we varied the
MF slope in the range $-1.0\div1.35$, searching the model that best
fits the radial density profile of the cluster.  Then we calculated
the radial variation of $x$ caused by mass segregation in the same
mass range as that of the observed stars.  The profiles are shown in
Fig.~\ref{pryor}.

It has been necessary to use anisotropy to fit the shallow external
part of the density profile of this cluster.  This means that the
stellar orbits of the cluster halo are preferentially radial,
beginning from an anisotropy radius $r_a$: after some iterations, we
found that $r_a = 25$ core radii provides the best fit.  Dynamically,
this is justified by the fact that the relaxation time in the halo of
this cluster exceeds the Hubble time, thereby implying that the stellar orbits
are not fully mixed (see Meylan \& Heggie 1997).  We have also
verified that isotropic models give approximately the same results as
reported here, even if their fitting of the radial profile is not as
satisfactory as the anisotropic one.

As expected, it can be seen in Fig.~\ref{pryor} that the MF slopes
increase from the inner to the outer bins.  Our two values of the
slope of the MF are consistent with the mass segregation of the
multi-mass King models, and the global MF slope turns out to be
$x_0=0.2\pm 0.3$.

\section{Summary and discussion} \label{thediscussion}

We have presented new large-field CCD photometry for $\sim20500$ stars in
the Galactic halo globular cluster NGC~1851, from both groundbased
observations and a pre-repair {\it HST} field. The photometric catalogue has
been used to build a $V$ vs. ($V$--$I$) CMD, which has been analysed in
detail. An extensive comparison of our data set with the predictions of the
stellar models has also been performed.

The effects of the dynamical evolution over the main sequence mass function
have been investigated by means of a completeness-corrected luminosity
function and the radial-count profile.

\paragraph{The evolved stellar content}

With an accurate inspection of the cluster bright-blue objects, and a
comparison with the numbers predicted from the background field and the
Galactic count models, we have confirmed the presence of seven ``supra-HB''
stars in the CMD of NGC 1851. We have shown that six of the ``supra-HB'' stars
could be evolved descendants from HB progenitors (post-HB or planetary
nebulae).

We have shown that standard evolutionary models are not able to reproduce
the observed bimodal distribution of stars along the HB.  Synthetic HR
diagrams demonstrate that the problem could be solved by assuming that the
efficiency of the RGB mass loss actually encompasses values going from 0.25
to 0.48.  We have found evidence that the radial distribution of the blue
HB stars is different from that of red HB and SGB stars.  The BHB stars are
significantly more concentrated than the SGB stars for $r>100$
arcsec. Though this distribution cannot be easily interpreted in terms of
dynamical evolution, it might be related to the anomalous distribution of
the BSs (see below).

All the 27 known variable stars have been identified, and 26 have been
measured in both colours (the remaining one being saturated). Twenty-two of
them are RR~Lyr variables. For the first time, our photometry has allowed
the mean absolute $I$ magnitude of the RR~Lyr variables to be obtained at a
metallicity [Fe/H]~=~--1.28. The RR~Lyr are brighter than the ZAHB in the
$V$ band, in accordance with the relation given by Carney et al. (1992).
The positions and the photometry for seven new RR~Lyr candidates have been
given. With these additional variables the ratio of the two types is now
$N_c$/$N_{ab} = 0.38$, which reduces the current estimate N$_c$/$N_{ab} =
0.47$ (Wehlau et al. 1982).

Thirteen BS stars have been identified outside the inner 80 arcsec. They do
not show any sign of variability. We have investigated the radial
distribution of the BSS. For $r>80$ arcsec, the BSs are less concentrated
than the SGB stars with the same $V$ magnitude. We argue that the
distribution of the BSs in the outer envelope of NGC 1851 might be similar
to the distribution found by Ferraro et al. (1997) for the BSs in the
envelope of M3.

We have considered a sample of 25 globular clusters and have derived a new
calibration for the $\Delta V_{\rm bump}^{\rm HB}$ parameter as a function
of cluster metallicity, and we have found that NGC~1851 follows this
general trend fairly well.  From a comparison with the corresponding slopes
predicted by the isochrones library from Bertelli et al.  (1994), we have
found that perhaps an age--metallicity relation actually exists among
globular clusters, with the metal poorest possibly being older.

\paragraph{Dynamical status of NGC~1851}

We have been able to derive a complete LF down to $V \simeq 23.5$ mag for
stars in the region 190\arcsec~ $< r <$ 650\arcsec~, and down to $V \simeq
22$ mag in the region 120\arcsec~ $< r <$ 189\arcsec~. The external LF is
steeper than the internal one, and we have interpreted this result as a sign
of mass segregation. By using the most updated mass--luminosity relations we
have obtained MFs which can be well fitted by power laws with distinct
exponents $x$. The observed value for the external MF is $x = 1.52 \pm
0.18$, which is steeper than the value $0.89 \pm 0.20$ found for the
internal one.

The global MF has been determined correcting the two observed mass functions
for the effects of mass segregation, as predicted by the multi-mass
King--Michie model which best fits the observed light profile of NGC 1851.
The two values for the slope of the MF are compatible with the model if a
global MF exponent $x_0=0.2\pm 0.3$ is adopted. This value for the global MF
slope is marginally smaller (MF flatter) than what would be expected from
the relation between the slope of the MFs and the position in the Galaxy and
the metallicity of the cluster proposed by Djorgovski et al. (1993). This
might indicate that NGC 1851 has had a stronger gravitational interaction
with the Galactic disc than the average of the Galactic GCs with similar
position and metallicity.

\paragraph{}
The above results indicate that NGC~1851 is a cluster where the dynamical
evolution has affected both its evolved and unevolved stellar content. While
the single findings are not of high statistical significance (mostly due to
the small size of the stellar samples), taken together they give a
coherent picture. Stellar encounters have led to mass segregation, as shown
by the MF, which is steeper and steeper going from external to internal
regions. They have probably contributed to the creation of the observed
group of blue straggler stars, and possibly have triggered the formation of
a blue tail in the HB.

The internal dynamics of NGC~1851 has therefore influenced the evolution of
its stars, introducing effects not reproducible by standard models.  In
turn, the dynamical evolution induced by the external gravitational field of
the Galaxy has also very probably contributed to the modification of the
present-day stellar population of NGC 1851, as strongly suggested by the
anomalously flat global mass function.

\begin{acknowledgements}
IS wishes to thank the {\em Instituto de Astrof\'{\i}sica de Canarias} for
providing a nice work environment during the reduction stage, and the {\it
Fondazione A. Gini} for partial financial support.  M. Catelan is warmly
acknowledged for comments that improved many of the points discussed. The
authors are also grateful to the anonymous referee for suggestions that
improved the original manuscript, and to Terry Mahoney for a careful
revision of the english text.  This project has been partially supported by
Italy's {\it Ministero della Ricerca Scientifica e Tecnologica} and Spain's
{\it Ministerio de Educacion y Ciencia}.
\end{acknowledgements}

\end{document}